\documentclass[10pt,conference,letterpaper]{IEEEtran}
\IEEEoverridecommandlockouts
\usepackage{cite}
\usepackage{amsmath,amssymb,amsfonts}
\usepackage{algorithmic}
\usepackage{bm}
\usepackage{graphicx}
\usepackage[bottom=0.9in,top=0.8in,left=0.68in,right=0.68in]{geometry}
\usepackage{textcomp}
\usepackage{xcolor}
\usepackage{ctable}
\usepackage{makecell}
\usepackage{multirow}
\usepackage{multicol}
\usepackage{booktabs}
\usepackage{tabularx}
\usepackage{subfigure}
\usepackage{url} 
\usepackage{verbatim}
\usepackage{algorithm}
\usepackage{balance}
\usepackage{titlesec}
\usepackage{fancyhdr}
\usepackage{balance}

\titlespacing{\subsection}{0pt}{1.5ex plus .2ex minus .2ex}{1ex plus .2ex}
\titlespacing{\section}{0pt}{1.5ex plus .2ex minus .2ex}{1ex plus .2ex}

\def\BibTeX{{\rm B\kern-.05em{\sc i\kern-.025em b}\kern-.08em
    T\kern-.1667em\lower.7ex\hbox{E}\kern-.125emX}}

\begin{document}

\title{JCSP: Joint Caching and Service Placement for Edge Computing Systems\\
\thanks{The work of Yicheng Gao has been supported by a China Scholarship Council-Imperial Scholarship. } \thanks{978-1-6654-6824-4/22/\$31.00 ©2022 IEEE}
}

\author{\IEEEauthorblockN{Yicheng Gao, Giuliano Casale}
\IEEEauthorblockA{Department of Computing, Imperial College London, London, UK \\
Emails: \{y.gao20, g.casale\}@imperial.ac.uk}}

\maketitle

\begin{abstract}
With constrained resources, what, where, and how to cache at the edge is one of the key challenges for edge computing systems. The cached items include not only the application data contents but also the local caching of edge services that handle incoming requests. However, current systems separate the contents and services without considering the latency interplay of caching and queueing. Therefore, in this paper, we propose a novel class of stochastic models that enable the optimization of content caching and service placement decisions jointly. We first explain how to apply layered queueing networks (LQNs) models for edge service placement and show that combining this with genetic algorithms provides higher accuracy in resource allocation than an established baseline. Next, we extend LQNs with caching components to establish a joint modeling method for content caching and service placement (JCSP) and present analytical methods to analyze the resulting model. Finally, we simulate real-world Azure traces to evaluate the JCSP method and find that JCSP achieves up to 35\% improvement in response time and 500MB reduction in memory usage than baseline heuristics for edge caching resource allocation.
\end{abstract}

\begin{IEEEkeywords}
Caching, service placement, queueing modeling, edge computing
\end{IEEEkeywords}

\section{Introduction}
Many types of smart applications for end users, such as virtual reality, interactive gaming, and face recognition, have rapidly evolved over the past decade \cite{mobiledevice}. Such compute-intensive applications are constantly expected to fulfill high data rates and meet quality of service (QoS) requirements arising from ever-increasing user demands.
However, the efficiency and reliability of these applications are both constrained by the terminal devices in terms of storage, battery life, and computation capability \cite{constrain}. Thus, to avoid high latency when processing incoming tasks, computationally-demanding jobs are typically offloaded to cloud data center back-ends for faster executions. 

However, this approach is faced with the issue of high response time, as the transmission distance from the end user to the data center can be significant \cite{continent}. This is problematic for latency-sensitive applications, such as autonomous driving and real-time video. In this context, a paradigm shift has taken place naturally from centralized cloud computing toward distributed edge computing \cite{emergence}. Edge computing configures computation, storage, and bandwidth resources at network edges, \emph{e.g.,} base stations and access points, close to end subscribers. Through these means, edge computing offers a dramatic reduction in latency. 

In this paper, we focus on edge caching strategies, \emph{i.e.,} what, where, and how to cache at the edge with limited resources to satisfy user demands, which are one of the key enabling technologies of edge computing. Traditional caches store copies of items, such as images, files, and scripts, at proxy servers or servers close to end users \cite{codedcache}. Different from traditional caching, edge caching stores items at proximal network edges, such as small cell base stations (SBSs) and vehicles. Moreover, the cached items are no longer limited to data contents but may also be seen as encompassing the decision of which services are deployed for execution at a particular edge resource. Specifically, content caching is proposed to tackle the issues of content placement and delivery \cite{replacement}. While service placement refers to the deployment of services and their correlative databases at edge servers to execute tasks offloaded from end users \cite{servicecache}. 

In existing studies, edge content caching specializes in data caching without considering queueing contention at the services that are placed to process it. In a similar manner, edge service placement typically overlooks the role content caching plays in the service execution latency. To tackle both issues, we propose a joint caching and queueing methodology consisting of a novel class of models and a resulting resource allocation approach. The unique feature of the models is the ability to stochastically model the latency interplay of queueing for service access and cache hits/misses. To implement this novel class of stochastic models, we generalize a class of models, called layered queueing networks (LQNs)~\cite{franktse}, which are used in the performance engineering of layered service architectures. Our extension integrates for the first time caching characteristics in these models, thus developing a novel tool for stochastic scheduling in edge systems. In details, our main contributions are as follows:

\begin{itemize}
\item We first model the basic job scheduling and service placement processes (ignoring caching) by LQNs. As applied to stochastic workloads, we find an average gain of 16$\%$ for system response time in our experiments compared to the deterministic scheduling method in \cite{dependenttask}. This illustrates the benefit of adapting an LQN formalism to deal with stochastic scheduling.

\item We generalize LQN models with caching components, which establishes a joint modeling method for edge content caching and service placement (JCSP), and present the analytical algorithm to analyze the generalized model numerically. 

\item We simulate an advanced Queueing Petri Net (QPN) based model to validate the accuracy of the proposed class of integrated LQN-caching models\footnote{The validation datasets are available at \url{https://doi.org/10.5281/zenodo.6491327 }} and implement the generalized LQN model within the LINE\footnote{The LINE tool is available at \url{http://line-solver.sourceforge.net/}} tool \cite{line}.

\item Extensive simulations based on real-world Azure traces are conducted to evaluate the applicability of our proposed JCSP method. Results show that the JCSP method can find a better trade-off between system response time and memory consumption under a wide range of situations than the baselines.
\end{itemize}

The paper is organized as follows. In Section II, we present related works and state the limitations. The methodology is developed in Section III. We design the generalized model containing caching and present the analytical methods in Section IV. The model validation and experimental evaluation are given in Section V. Finally, Section VI concludes the paper.

\section{Related Work}
Current research has investigated edge caching from two perspectives, \emph{i.e.,} content caching and service placement, as shown in Table \ref{literature}. The basic issue for edge content caching is optimizing what to cache and how, based on the architectural features of the edge computing system under study. With respect to what to cache, most studies center on the characteristics of content popularity profile, which can be grouped into known \cite{knownpopu}, unknown but static \cite{constpopu1}, and unknown and time-varying \cite{varypopu1,varypopu2}. With regard to how to cache instead, policies differ in the setting of network topologies. When employing traditional two-tier hierarchical networks, the key issue is to make a trade-off between the edge node and the centralized node under certain resource constraints \cite{edgecachingicc}. When adopting the user-centric structures, how to collaborate within a cluster of edge nodes, such as SBSs or vehicles, is crucial \cite{usercentric,proactivecache}. The aforementioned studies on caching contents at edge nodes ignore the services that are placed to process such contents.

\renewcommand\arraystretch{1.2} 
\begin{table}[t]\scriptsize
\centering
\setlength{\belowcaptionskip}{10pt}
\caption{Comparison of related works with different parameters}
\scalebox{0.9}{
\begin{tabular}{
m{1.1cm}<{\centering} 
m{0.6cm}<{\centering} 
m{0.6cm}<{\centering} 
m{1.2cm}<{\centering} 
m{1.0cm}<{\centering}
m{1.0cm}<{\centering}
m{1.2cm}<{\centering}}
  \toprule
  \multirow{2}*{Ref. }
  &\multicolumn{2}{c}{Objective items} 
  &\multirow{2}*{\makecell[c]{Replacement\\ policies}}
  &\multicolumn{2}{c}{Job scheduling}
  &\multirow{2}*{\makecell[c]{Job\\ precedences}} \\
  \cline{2-3}  
  \cline{5-6}  
  & content & service &   & deterministic & stochastic &
  \\ \hline
  
  \multirow{1}*{\cite{knownpopu,constpopu1,varypopu1,varypopu2}}
  & $\surd$
  & 
  & 
  & 
  & 
  &  \\ \hline
  
  \cite{edgecachingicc},\cite{proactivecache}
  & $\surd$
  & 
  & $\surd$
  & 
  & 
  &  \\ \hline
  
  \multirow{1}*{\cite{vr,availability,collaborate}}
  & 
  & $\surd$
  & 
  & 
  & 
  &  \\ \hline

  \cite{loaddispatching}
  & 
  & $\surd$
  & 
  & $\surd$
  & 
  &  \\ \hline
  
  \cite{servicecach1,servicecach2}
  & 
  & $\surd$
  & 
  & 
  & $\surd$
  &  \\ \hline

  \cite{functionconfig,dependenttask}
  & 
  & $\surd$
  & 
  & $\surd$
  & 
  & $\surd$\\ \hline
  
  \textbf{JCSP}
  & $\surd$
  & $\surd$
  & $\surd$
  & 
  & $\surd$
  & $\surd$\\ 
  \bottomrule
  
\end{tabular}}
\label{literature}
\end{table}

For edge service placement, existing studies center on optimizing the system performance \cite{vr,availability,collaborate} , \emph{e.g.,} response time, energy efficiency, or cost, due to the constrained edge resources. 
Further, many studies jointly optimize service placement with job scheduling processes \cite{loaddispatching,servicecach1,servicecach2}. The two procedures are interactive, as the placed services determine how to schedule the jobs and the scheduling strategy impacts the performance of the placement policy. To jointly model the two procedures, most techniques take advantage of queueing theory, such as $M/G/1$\cite{servicecach1} and $M/M/1$\cite{servicecach2}. However, these models typically do not reflect the precedences between jobs. If considering the precedence constraints, most approaches adopt directed acyclic graphs (DAG) \cite{functionconfig,dependenttask}, the optimal scheduling of which is typically NP-hard except with specific graph topologies (\emph{e.g.,} in-trees, out-trees).

\begin{figure} [t]
\centering
\includegraphics[width=0.48\textwidth]{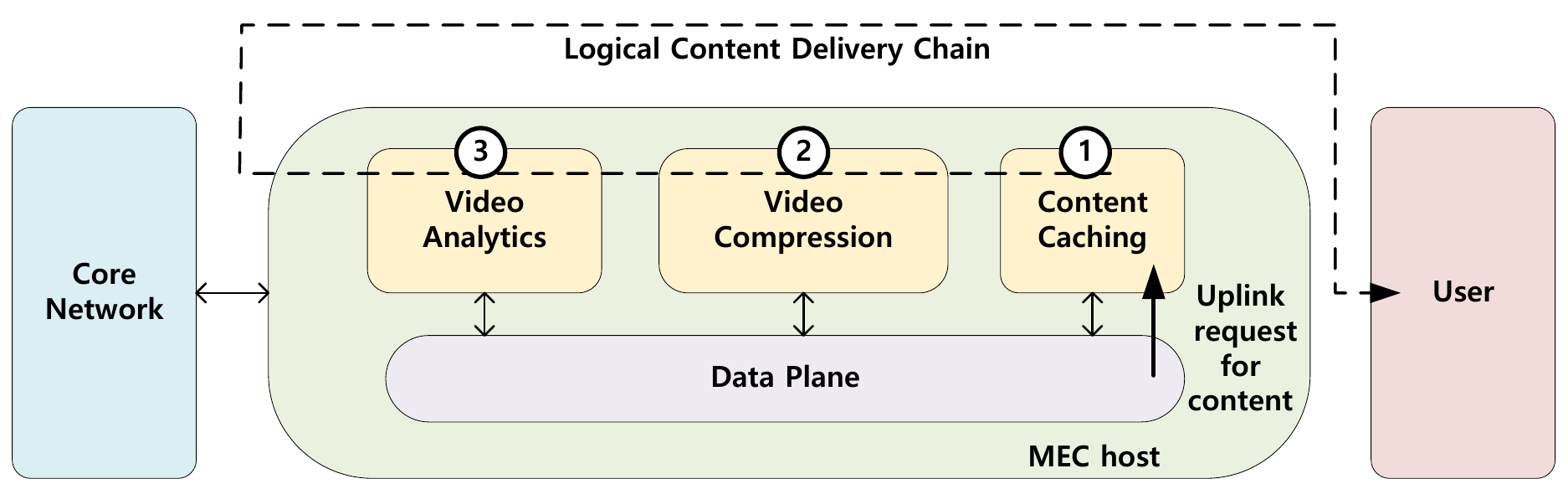}
\caption{A use case of video content delivery chain}
\label{casestudy-video}
\vspace{-0.4cm}
\end{figure}

In the above-mentioned research, there are two primary limitations. First, for edge service placement, DAGs are efficient to model the jobs with precedences. However, a DAG fails to illustrate the dependencies of underlying resources invoked by jobs, because it does not encompass scheduling strategies and resource contention. Therefore, we propose to jointly model the dependent job scheduling and service placement processes by LQNs. LQNs have been extensively used to model distributed computing systems with complex characteristics, such as fork/join interactions and multiple tiers of services \cite{lqn}. An LQN distributes software and hardware resources into multiple layers, according to their functionality and request types. When there are requests need to possess simultaneous resources \cite{franktse}, LQNs can explicitly represent the dependencies by describing the interaction between clients and servers in each layer. Each layer is then automatically translated into a Markovian queueing network model, which is analyzed in isolation, prior to aggregating the results to produce end-to-end latency estimates. Thus, if applying LQNs to layered edge computing systems, each edge node can either act as a client or as a server to request or provide services, which makes it easy to model and trace nested dependencies.

Next, the edge content caching and service placement processes are studied separately. However, in practice, the processing of many applications requires both computation and data. For example, in video analytics \cite{etsi}, the request is first forwarded to the content caching function to retrieve the cached contents. After the identification, the request is routed to video compression and video analytics services for further processing, as shown in Fig. \ref{casestudy-video}. Therefore, we propose to establish a unified resource allocation approach for both contents and services. As we define caching and LQNs to model processing for contents and services respectively, the key challenge is how to establish and analyze integrated caching and queueing network models.

\section{Methodology}

\subsection{System Model}

\begin{figure} [t]
\centering
\includegraphics[width=0.48\textwidth]{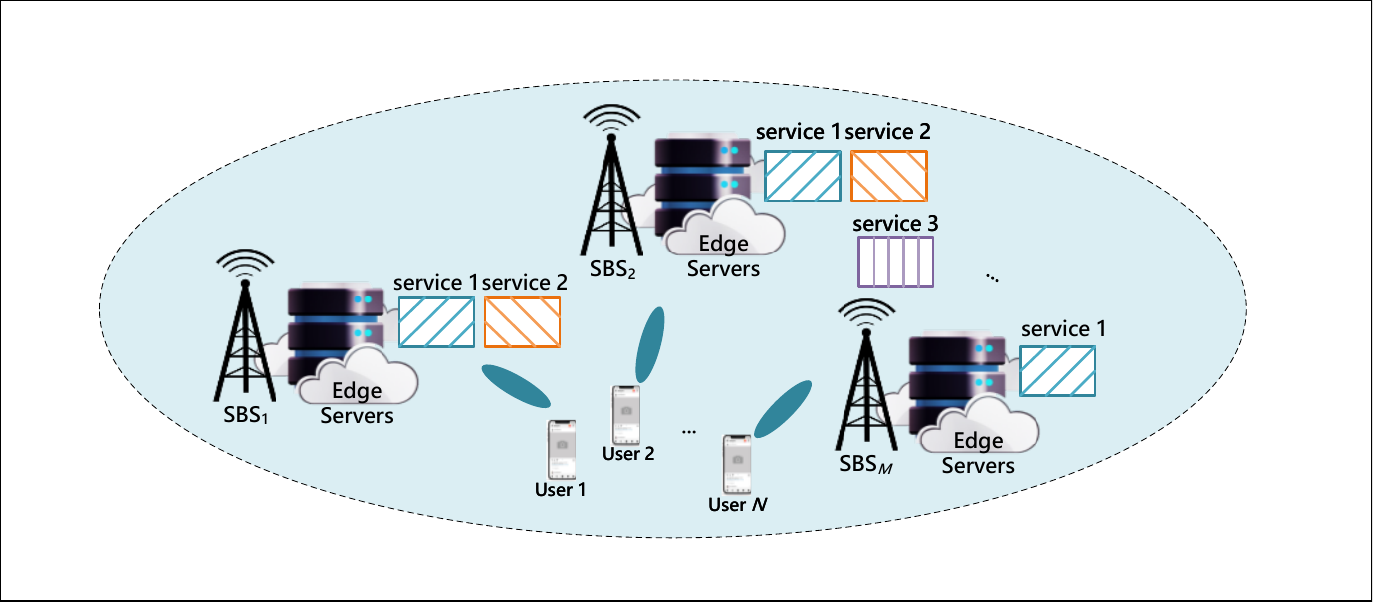}
\caption{Edge computing system model containing $N$ users}
\label{edgecachingsystem}
\vspace{-0.4cm}
\end{figure}

We consider a cluster of $M$ edge nodes connecting $N$ users in the edge computing system, as shown in Fig. \ref{edgecachingsystem}. Each edge node provides multiple services in edge servers co-located with the SBSs. Each user offloads several jobs to the connected edge node to be served. The offloaded jobs may have precedences that must be processed in order. For example, a face recognition application \cite{facerecognition} can be regarded as consisting of five successive jobs, \emph{i.e.,} object acquisition, face detection, pre-processing, feature extraction, and classification. Each job can be operated solely at the edge node that provides corresponding services. When the requested services are not provided in the connected edge node, the jobs need to be forwarded to other edge nodes that satisfy the requirements. Furthermore, different types of jobs at the same node also obey the sequential processing. 

We assume the total number of jobs is $K$, which request a total number of $C$ possible services. For job $k$ at node $m$, $k=1,2,...,K, m=1,2,...,M$, we define two parameters as follows:
\begin{itemize}
    \item \emph{{Service Placement Decision}} $p_{mk}$. If node $m$ provisions the service requested by job $k$, $p_{mk}=1$. Otherwise, $p_{mk}=0$. The set of all nodes that provision the service for job $k$ is denoted as $\bm{F}_k=\left\{m|p_{mk}=1, m=1,2,...,M\right\}$.
    \item \emph{{Mean service time}} $t_{mk}$. This is the amount of computational processing requested by job $k$ upon visiting node $m$. Different jobs may request different mean service time.
\end{itemize}

\begin{figure} [t]
\centerline{\includegraphics[width=0.48\textwidth]{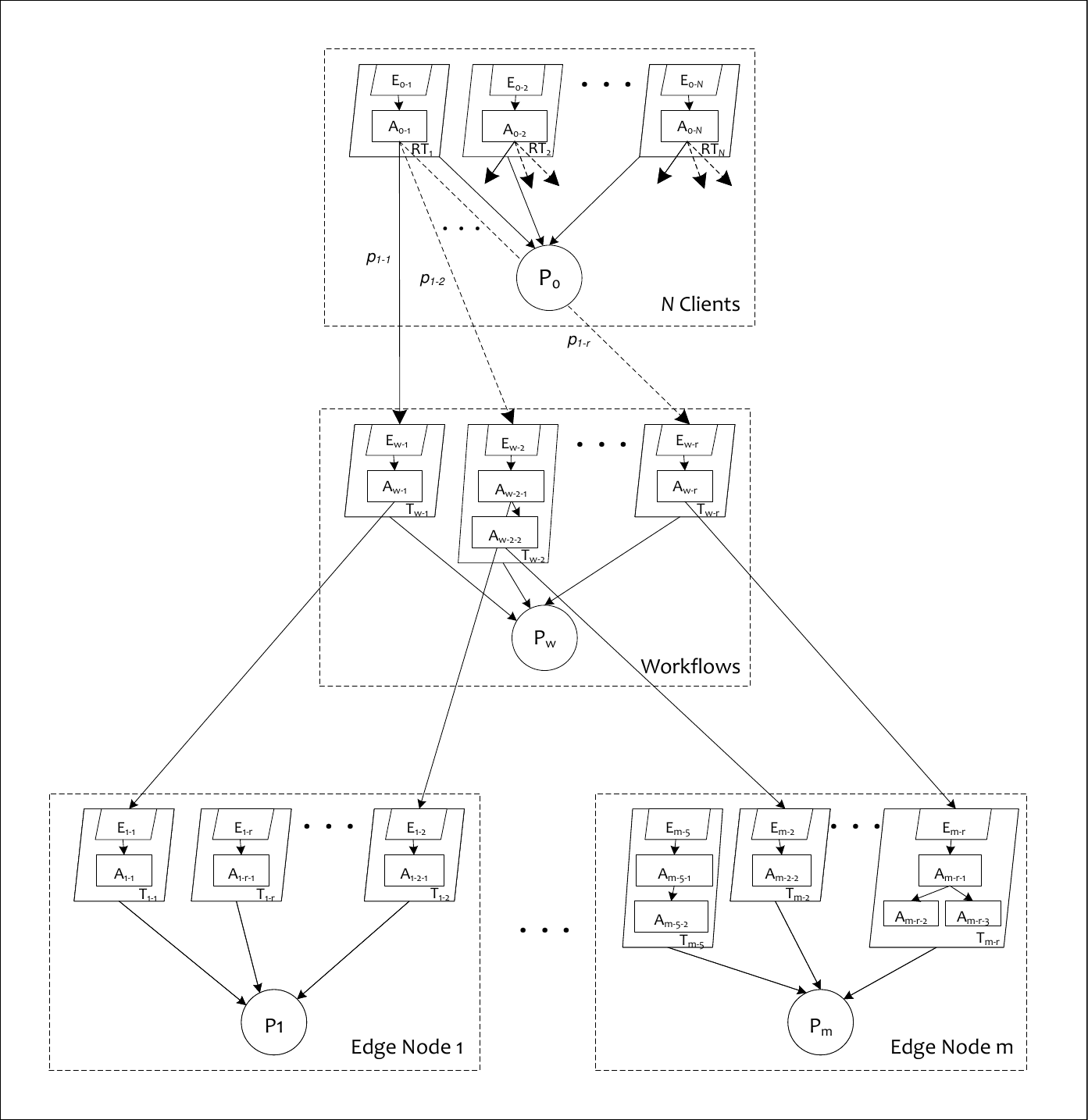}}
\caption{LQN-based model for job scheduling and service placement processes}
\label{lqnmodel}
\vspace{-0.4cm}
\end{figure}

We then use an LQN to model the job scheduling and service placement problem, as presented in Fig. \ref{lqnmodel}. Each edge node $m$ is composed of 4 components as follows:
\begin{itemize}
    \item \emph{{Processor}}: The SBS is modeled as a processor $P_m$, shown as a circle. It obeys the processor-sharing scheduling policy, which is a mathematical abstraction of time-sharing scheduling with an infinitesimally small quantum of time.
    \item \emph{{Task}}: The co-located edge server is modeled as a set of tasks $T_{mc}$, shown as the larger parallelograms, each of which represents service $c$ running on processor $P_m$. Each task has a queue to model first-come first-served (FCFS) admission control and the service time follows a phase-type distribution, \emph{e.g.,} exponential, Erlang, or hyper-exponential.
     \item \emph{{Entry}}: Each task publishes at least one entry $E_{mcj},1\leq j\leq J$, shown as a small parallelogram, which represents the end point for calls to service $c$. In general, a service with $k$ possible endpoints is modeled as a task with $k$ entries.
     \item \emph{{Activity}}: Each entry invokes a sequence of activities $A_{mci}$, where $i=1,\ldots,I_c$ and $I_c$ is the number of activities for service c. Activities are shown as a series of rectangles under the entries, to represent independent or dependent executions. That is, upon invoking an entry $E_{mcj}$, it will lead to a chain of individual activities being carried out on the underpinning processor. In LQNs, such activities can be organized in form of a DAG called the activity graph and detailed later. Note that each activity may be reached by calling different entries.
\end{itemize}

The clients are modeled in the LQN formalism as \emph{reference tasks} running on a \emph{pseudo-processor} $P_0$. A pseudo-processor is a modeling abstraction similar to a processor but without a physical reality in the edge system and used to keep a conceptually similar description of all tasks in the model. The pseudo-processor in LQNs is also suitable to describe user think times. Each client has different probabilities of requesting different workflows. The user workflows are also DAGs to represent the precedence between the jobs and modeled as special activity graphs within the {pseudo-processor} $P_w$. The mean queue length of the infinite server on $P_{w}$ is defined as the multiplicity of the workflow. If dependent jobs are executed at different nodes, the activity graph will feature special activities that issue synchronous calls to different nodes, such as $A_{w-2-1}$ and $A_{w-2-2}$. Therefore, DAGs are present both in the reference processor to describe user workflows and bound to entries of the services to describe the activities launched by each entry. The latter is also in general DAGs that may include synchronous or asynchronous calls to other tasks and entries.

\subsection{Problem Formulation}
The LQN models we have described above allow to obtain analytical estimates of throughputs, response times (\emph{i.e.,} latency per unit invocation), mean queue-lengths (\emph{i.e.,} backlogs), and utilizations for each component type. We focus on response time throughout the rest of the paper, but other reference metrics may be easily generated from the models. Our optimization objective is to minimize the end-to-end total response time $R$ of the LQN-based model, which is expressed as
\begin{equation}
R=\sum_{i=1}^C{\frac{X_i}{X}}{R_i},
\end{equation}
where ${X_i}$, ${{R_i}}$, and ${X}$ are the throughput of job class $i$, the response time of job class $i$, and the total throughput summed over all service classes, respectively. The ratio $X_i/X$ may be seen as related to the probability at steady-state where an arrival is of class $i$. Thus, the problem of dependent job scheduling and service placement is formulated as
\begin{equation}
    \begin{split}
        \min_{\bm{x}} \quad\quad& R(\bm{x}), \\
        \emph{s.t.} \quad\quad & \sum \limits_{m \in \bm{F}_k} x_{mk}=1, \\
        & x_{mk} \in \{0,1\}, 
    \end{split}
\end{equation}
where $x_{mk}$ denotes the scheduling decision for job $k$ at node $m$. The vector $\bm{x}=[x_{11},...,x_{mk},...,x_{MK}]$ represents the set of scheduling decisions for all jobs and $R(\bm{x})$ is the corresponding system response time with decision vector $\bm{x}$. The first constraint guarantees that job $k$ is solely served at one (and only one) feasible edge node provisioning the requested services. The second constraint ensures that if job $k$ is scheduled to be served at node $m$, then $x_{mk}=1$, or $x_{mk}=0$ otherwise.

\section{Generalized LQN Model Design}

\subsection{New Design Formalism}

To extend LQN models to include caching components, we first define two novel components as follows:
\begin{itemize}
\setlength{\itemsep}{0pt}
\item \emph{cache-task}: Each caching node is defined as a cache-task in the LQN model. Each cache-task offers the ability to access a collection of items through a cache, such as a key-value store. Cache-tasks have the basic properties of tasks, but add four specific properties for caching: the total number of items, the cache capacity, the cache partitioning into lists, and the cache replacement policy, \emph{e.g.,} least recently used (LRU), least frequently used (LFU), first-in first-out (FIFO), and random replacement (RR). 
\item \emph{item-entry}: The services provided by the caching module are defined as item-entries in the LQN model. Each item-entry represents a collection of items accessible through the cache-task. Item-entries are similar to standard entries, but add the property of the popularity of the items, \emph{e.g.,} Zipf or custom distribution. 
\end{itemize}

Upon issuing a call to an item-entry exposed by a cache-task, a client obtains the requested object either after a cache hit or a cache miss. The hit/miss selection depends on the cached contents and reflects as a choice of a different branch in the DAG bound to the item-entry. This is modeled through a novel precedence relationship, \emph{i.e.,} an activity graph edge. We name this precedence relationship as \emph{cache-access} for the cache hit and miss activities under each item-entry, which can be used to specify the branching in the activity graph. The \emph{cache-access} transforms an incoming job into a hit or miss job, sending out the transformed job to the corresponding outgoing link with a certain probability.

The cache replacement policy of each cache-task adopts by default the RR policy, which randomly selects a candidate item to replace when a cache miss occurs. The RR policy can offer the same steady-state performance of the time-to-live (TTL) mechanism \cite{ttl} with exponential timers. Hence it is a reasonable approximation of the real-world TTL with better analytical tractability. Other policies, such as LRU or CLIMB, can be obtained as simple extensions of our basic framework.

\begin{figure} [t]
\centering
\includegraphics[width=0.45\textwidth]{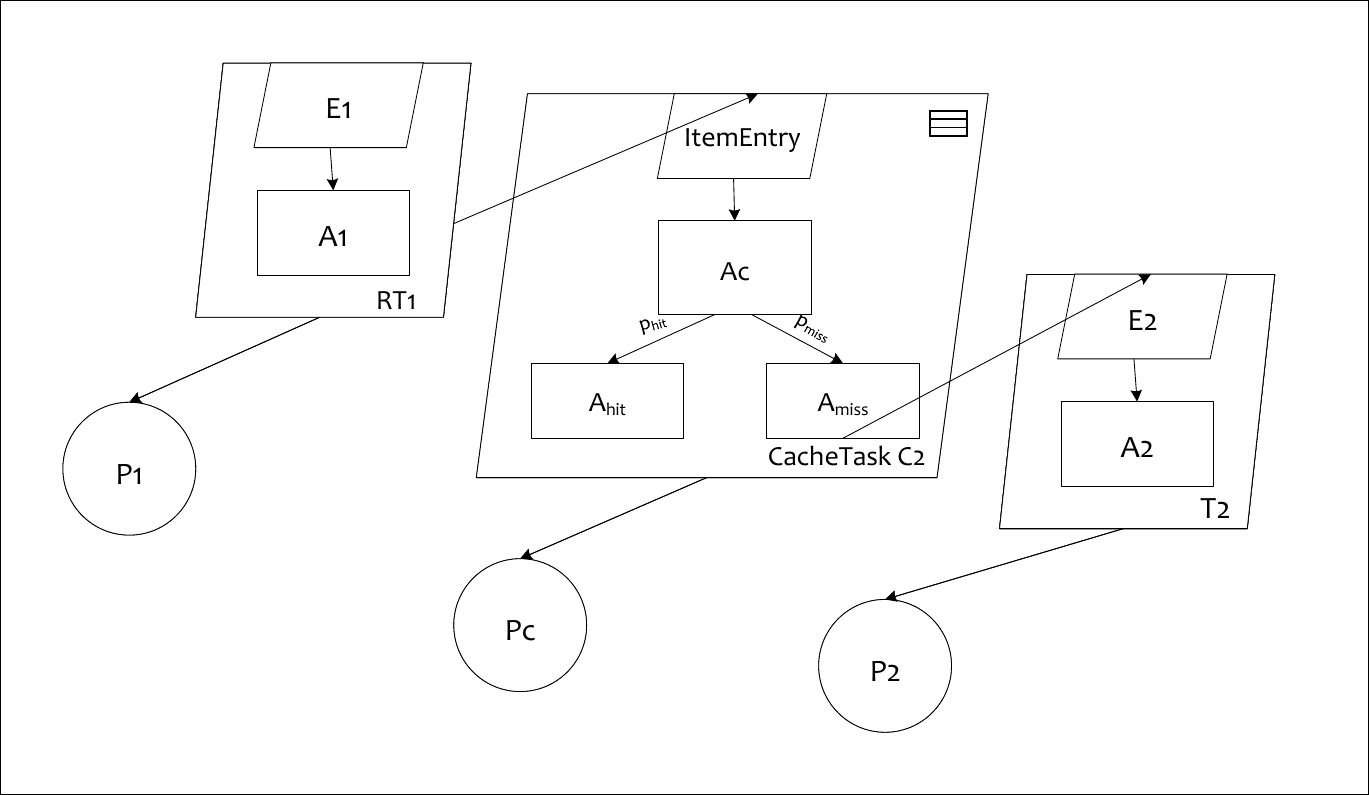}
\caption{An example of a LQN model containing one cache-task}
\label{cachetask}
\vspace{-0.4cm}
\end{figure}

Based on such extensions, we generalize the LQN model by enabling caching. Fig. \ref{cachetask} presents an example of an LQN model containing one cache-task. The top-right symbol in the task parallelogram distinguishes a cache-task from an ordinary task. 
When an item request arrives, we model the probability that the item is in the cache with $p_{hit}$ and out of cache with $p_{miss}=1-p_{hit}$.
The generated architecture allows joint modeling of caching and queueing into the same LQN model. It is important to note that $p_{hit}$ and $p_{miss}$ need {\em not} be specified by the end user. They are automatically determined through the solution of the stochastic models underpinning the LQNs with caching, which also relies on a stochastic equilibrium analysis of the cache replacement policy, as we describe in the next section.

\subsection{Analysis of LQN Models with Caching}

\begin{figure} [t]
\centering
\includegraphics[width=0.48\textwidth]{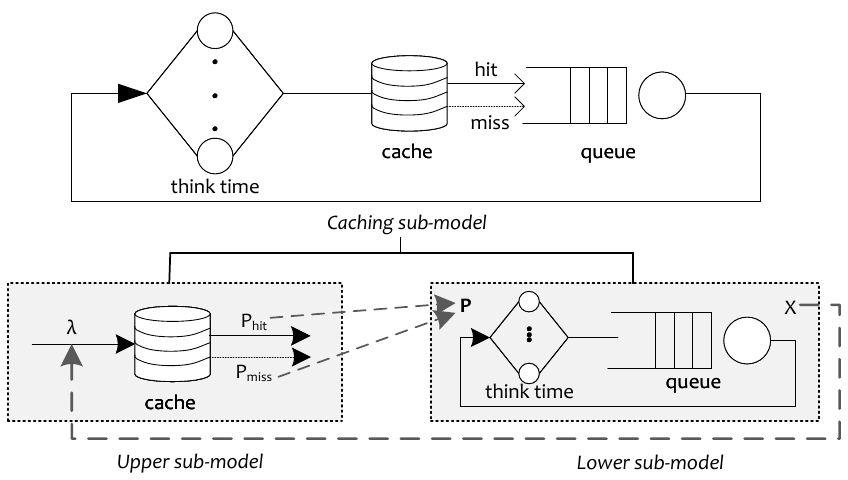}
\caption{Decomposition of the caching sub-model in the LQN model}
\label{cachesolver}
\vspace{-0.4cm}
\end{figure}

To solve the LQN model containing caching modules, the first step is to decompose the entire model into a group of sub-models. Each sub-model may be seen as a mixed queueing network that contains clients and servers. The term mixed refers to the simultaneous presence of open and closed service classes. The former represent asynchronous requests from clients, while the latter represent synchronous ones (\emph{i.e.,} closed since issued from a finite connection pool).
In particular, when a queueing network contains the cache module, the workload is synchronous, which means the issue of the next request to the cache is required to take place after the reply of the last request issued to the cache. To model this dynamics, we additionally divide the caching sub-model into two sub-models, as shown in Fig. \ref{cachesolver}. In the upper sub-model, the cache module is isolated in an open queueing network with asynchronous Poisson streams. While in the lower sub-model, the delay and the queueing station are contained in a closed queueing network.

To solve the caching sub-model, we generalize the analytical method for non-layered queueing networks from \cite{ton} to the multi-layer setting of LQNs. The expressions we give for the method focus for simplicity on a single class of jobs, but our implementation applies similar formulas developed in \cite{ton} for the multi-class case. 

Cache requests arrive in the upper sub-model as a single-class Poisson stream with rate $\lambda=\sum_r \lambda_r$. With an initial guess of the arrival rate $\lambda$, the hit and miss rates of the cache node, \emph{i.e.,} $p_{miss}^{(t)}$ and $p_{hit}^{(t)}$, are approximated iteratively via a fixed-point iteration (FPI) scheme. The obtained values in each iteration are leveraged in \cite{ton} to calculate the next arrival rate $\lambda^{(t+1)}$ by the following application of Little's law:
\begin{equation}
\label{eq:fpi}
\lambda^{(t+1)}=\frac{s\lambda^{(t)}}{\lambda^{(t)}\theta_t+p_{miss}^{(t)}\theta_m+p_{hit}^{(t)}\theta_h},
\end{equation}
where $\theta_t, \theta_m, \theta_h$ are the mean think time, and the mean delay to due miss and item fetch respectively. Here, $s$ is the maximum number of pending requests to the cache, \emph{i.e.,} the total population of jobs in the caching sub-model. The iteration process continues until the value of $\lambda$  converges. 

The hit and miss rates at each iteration are used to parameterize the queueing network of the \emph{lower} sub-model as routing probabilities. That is, jobs representing calls that experience cache hits will dynamically switch after visiting the cache to a \textit{hit service class}, which will be processed by the queueing station according to the activities specified in the activity graph for hits. Similarly, calls that experience misses will switch to a \textit{miss service class} representing the workflow in the activity graph for misses. This class-switching based separation of the hit and miss calls enables the modeling of the different response times experienced by the two classes of requests, which depend on whether the items they accessed were cached or not. At the end of the chain of operations that follow hit/miss, both classes of jobs will be merged back into the original class and return to the user as a response for that class.

The solution to the upper sub-model is reconciled with the lower sub-model iteratively by means of the following scheme:
\begin{itemize}
\setlength{\itemsep}{0pt}
\item Guess $\lambda^0$ with a random value.
\item Solve the upper sub-model with approximation methods, \emph{e.g.,} FPI, to obtain the hit and miss probabilities, \emph{i.e.,} $p_{hit}$ and $p_{miss}$.
\item Pass $p_{hit}$ and $p_{miss}$ to the lower sub-model to set the routing matrix.
\item Solve the lower sub-model with a queueing network solver method $f(s)$, \emph{e.g.,} approximate mean value analysis (AMVA) \cite{bolch}, to obtain the mean throughput $X$.
\item Pass $X$ to the upper sub-model as the next arrival rate $\lambda^n$.
\item Update until $||\lambda^{n+1}-\lambda^n||<\delta$.
\end{itemize}
We have never observed situations where the above iteration fails to converge. Based on this solution to the caching sub-model, the entire procedure to solve the LQN model containing cache-tasks is presented in Algorithm \ref{solver}, where $f(\cdot)$ is the customized solver for the sub-model $s$ that returns throughputs, queue-lengths, response times, and utilizations.

\begin{algorithm}
\caption{Solution algorithm for LQNs with caching}\label{solver}
\textbf{Input:} LQN model with cache-tasks, Sub-model solution algorithm $f(\cdot)$ 
\begin{algorithmic}[1]
\STATE Decompose $S$ sub-models based on layers 
\WHILE{not converged or not reached iteration limit}
\FOR{$s \gets 1$ to $S$}
\IF{$s \leftrightarrow$ caching sub-model}
\STATE Decompose sub-model $s$ into two sub-models.
\STATE Initialize $\lambda$ in the upper sub-model.
\WHILE{$||\lambda^{n+1}-\lambda^n||\geq \delta$}
    \STATE Solve the upper sub-model with \eqref{eq:fpi} for $\lambda$.
    \STATE Obtain $p_{hit}$ and $p_{miss}$ using the FPI method. 
    \STATE Pass $p_{hit}$ and $p_{miss}$ to the lower sub-model. 
    \STATE Solve the lower sub-model with AMVA.
    \STATE Return the throughput $X$ to the upper sub-model as the next arrival rate $\lambda^n$.
\ENDWHILE  
\ELSE
    \STATE Solve sub-model $s$ by $f(s)$.
    \STATE Set waiting time for sub-model $s$.
    \STATE Set think time for sub-model $(s + 1)$.
  
\ENDIF
\ENDFOR
\ENDWHILE 
\end{algorithmic}
\textbf{Output:} layer response times, residence times, throughputs, and utilizations.
\end{algorithm}

\subsection{Cache parameters and Markov process}

The caching model mentioned above for analyzing cache hit and miss probabilities applies list-based caches \cite{ton,gast}. The list-based cache consists of $h$ lists formed in a tree topology, each of which has a capacity of $m_l$ items, $l=1,2,...,h$. The total cache capacity is $m=\sum_{l=1}^h m_l$, which does not exceed the total number of items $n$. The items that are not cached are arranged in a virtual list with $l=0$, with a capacity of $m_0=n-m$. The request rate $\lambda_{vk}(l)\geq 0$ issued by stream $v$ for item $k$ in list $l$ follows a Poisson arrival process, where $v=1,2,...,u$, $k=1,2,...,n$.  

Each list can only have one parent list $p(l)$ to exchange the items, satisfying $p(p(\cdot\cdot\cdot p(l)\cdot\cdot\cdot))=0$, but has no constraint to the number of children lists. Lists without children are known as leaf lists. The probability of item $k$ shifting from current list $l$ to the list $j$ after a cache hit is non-uniform and defined as the access probability $c_{vk}(l,j)$. The analytical model for such list-based caches is based on a Markov process. In each list, the state vector $\bm{s}=[s(i,j)] \in [1,n]$ represents the item cached in position $i$ of list $j$. The applied replacement policies are modeled as a continuous-time Markov chain (CTMC) with an equilibrium probability $\pi(\bm{s})$. The product-form solution to $\pi(\bm{s})$ is presented as
\begin{equation}
    \pi(\bm{s})=\frac{\prod \limits_{j=0}^{h}\prod \limits_{i=1}^{m_j} \gamma_{s(i,j)j}}{E(\bm{m})},
\end{equation}
where ${E(\bm{m})}$ is the normalizing constant and $\gamma_{ij}$ denotes the access factor of item $i$ to access list $j$, satisfying
\begin{equation}
    \gamma_{ij}=\gamma_{ip(j)}\sum \limits_{v=1}^{u} \lambda_{vi}(p(j))c_{vi}(p(j),j).
\end{equation}

The performance measure for the Markovian analytical model is defined based on the marginal probability of item $k$ in list $j$ as
\begin{equation}
    \pi_{kl}(\bm{m})=m_l\gamma_{kl}\frac{E_k(\bm{m}-1_l)}{E(\bm{m})}.
\end{equation}
$\pi_{k0}(\bm{m})$ represents the miss ratio for item $k$, satisfying $\pi_{k0}(\bm{m})=E_k(\bm{m})/E(\bm{m})$, in which $E_k(\bm{m})$ is the normalizing constant when item $k$ is not in the model. From the marginal probabilities, the miss rate for requesting item $k$ from stream $v$ can be denoted as $M_{vk}(\bm{m})=\lambda_{vk}(0)\pi_{k0}(\bm{m})(1-c_{vk}(0,0))$. Thus, the total cache miss rate is $M(\bm{m})=\sum_{v,k}M_{vk}(\bm{m})$. Throughout, we focus on single-list caches, \emph{i.e.,} $h=1$, as they are more widespread, but our implementation also supports $h>1$ through the formulas above.

\section{Experimental Evaluation}
In this section, we first evaluate the merits of utilizing LQNs to model the job scheduling and service placement problem. Then we simulate a QPN-based model with Java Modeling Tools (JMT) \cite{jmt} to validate the accuracy of the generalized LQN model. Finally, we conduct extensive simulations of real-world Azure traces to measure the performance of the generalized LQN model.

\subsection{Evaluation of LQNs for Edge Service Placement}
We begin by comparing our proposed scheme with the method in \cite{dependenttask}, which formulates an offloading dependent tasks with service caching (ODT-SC) problem. The ODT-SC problem considers deterministic scheduling with deterministic service time, while our scheme focuses on stochastic scheduling with random service time. Moreover, the optimization goal of the ODT-SC problem is to minimize the maximum makespan, \emph{i.e.,} the sum of the start time of a job and its execution time, while our goal is to minimize the total response time. Despite the differences in the assumptions, we regard this baseline as a suitable comparison to just the benefits of stochastic LQN-based modeling of complex edge workflows. 

\begin{figure} [t]
\centering
\includegraphics[width=0.48\textwidth]{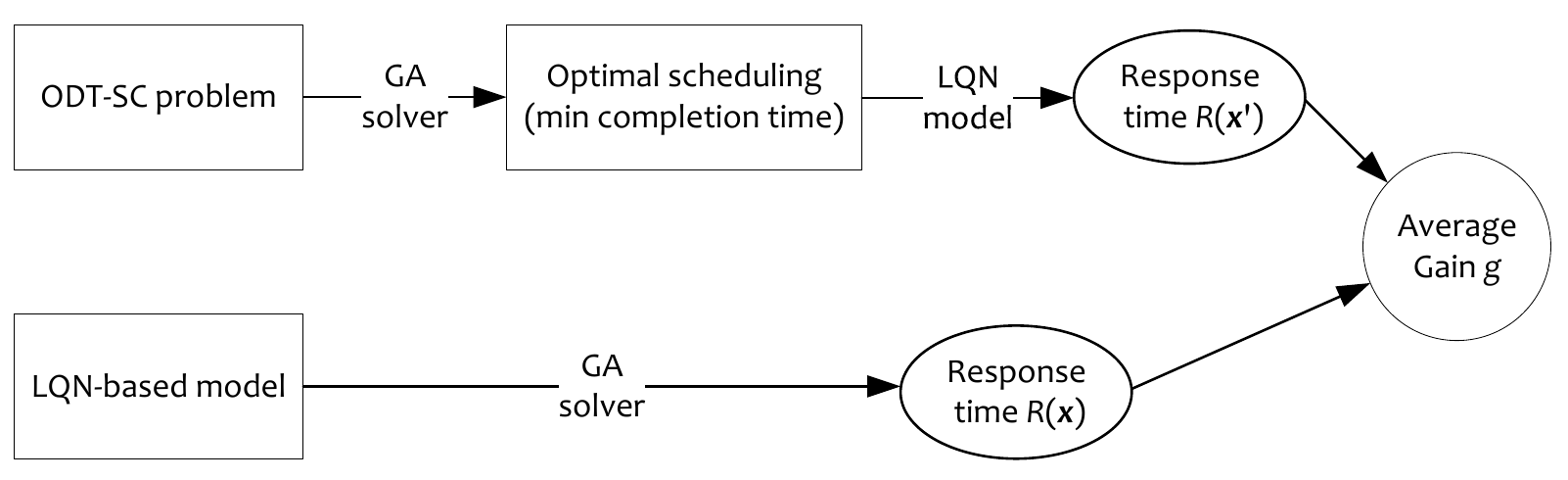}
\caption{An overview of the comparative procedure between the LQN-based model and the method in \cite{dependenttask} }
\label{compareflow}
\vspace{-0.4cm}
\end{figure}

For comparison, we set the mean service time for each job in our LQN-based model equal to the execution time in \cite{dependenttask}. The optimization goal of the ODT-SC problem is reset to minimize the total completion time, which is the sum of the makespan of each job. Moreover, the method in \cite{dependenttask} reformulates the ODT-SC problem into a convex programming to get feasible solutions. To be more general, we solve the ODT-SC problem by genetic algorithm (GA) without the need to relax the constraints. An overview of the comparative procedure of the two schemes is presented in Fig. \ref{compareflow}.

We first leverage the GA to address the ODT-SC issue, obtaining the optimal scheduling strategy $\bm{x'}$of each job. Then the strategy is transformed into an LQN model to calculate the total response time $R(\bm{x'})$. The response time corresponding to the minimum completion time in each iteration is selected as the comparative object. Meanwhile, we equally exploit the GA to solve the proposed LQN-based model to search for the minimum total response time. Finally, we compare the two response times to observe the average gain $g$ of the proposed method, which is expressed as
\begin{equation}
g=\frac{1}{I}\sum_{i=1}^I\frac{\parallel R_i(\bm{x'})-{{R_i(\bm{x})}}\parallel}{{R_i(\bm{x'})}} ,
\end{equation}
where $I$ is the total iterations.

In each iteration $i=1,\ldots,I$, to reduce the estimation error, we run $Q$ replications of the GA solver to search for the optimal solutions. In each replication of the GA solver, the number of generations is denoted by $G$. The dimension of the design variable for the ODT-SC problem is $2K$, of which the first and second $K$ elements are the start times and scheduling decisions of each job respectively. The service time is deterministic but varying in different kinds of services. While for the LQN-based model, the dimension of the design variable is the $K$ scheduling decisions of each job. The service time of each job class is set to be exponentially distributed or hyper-exponentially distributed, of which the mean value is equal to the one in the ODT-SC problem.

A set of experiments are conducted to analyze the performance under different conditions. The setting of different parameters are shown in Table \ref{exparameters}. The distribution of service time is assumed as exponential or hyper-exponential. For simplicity, the user workflow adopts chain structures for multiple services in the experiments, but it is easily extended to cases with parallel structures by LQNs. 

\begin{table}[t]
\setlength{\belowcaptionskip}{0.5cm}
\centering
\caption{Parameter settings for the comparative experiments} 
\renewcommand\arraystretch{1.25}
{\begin{tabular}{lll  }
\toprule 
\textbf{Definition} &\textbf{Parameter} &\textbf{Range of Value} \\
\hline
number of edge nodes& $M$& $2,5$  \\
number of users &$N$ & $8,20$\\
number of services &$C$ & $3,9,15$  \\
number of iterations / replications&$I$ / $Q$ & $30$  \\
number of generations &$G$ & $1000$  \\
\bottomrule
\end{tabular}}
\label{exparameters}
\end{table}

\begin{figure} [t]
\centering
\subfigure[Exponential Service Time]{
\includegraphics[width=0.23\textwidth]{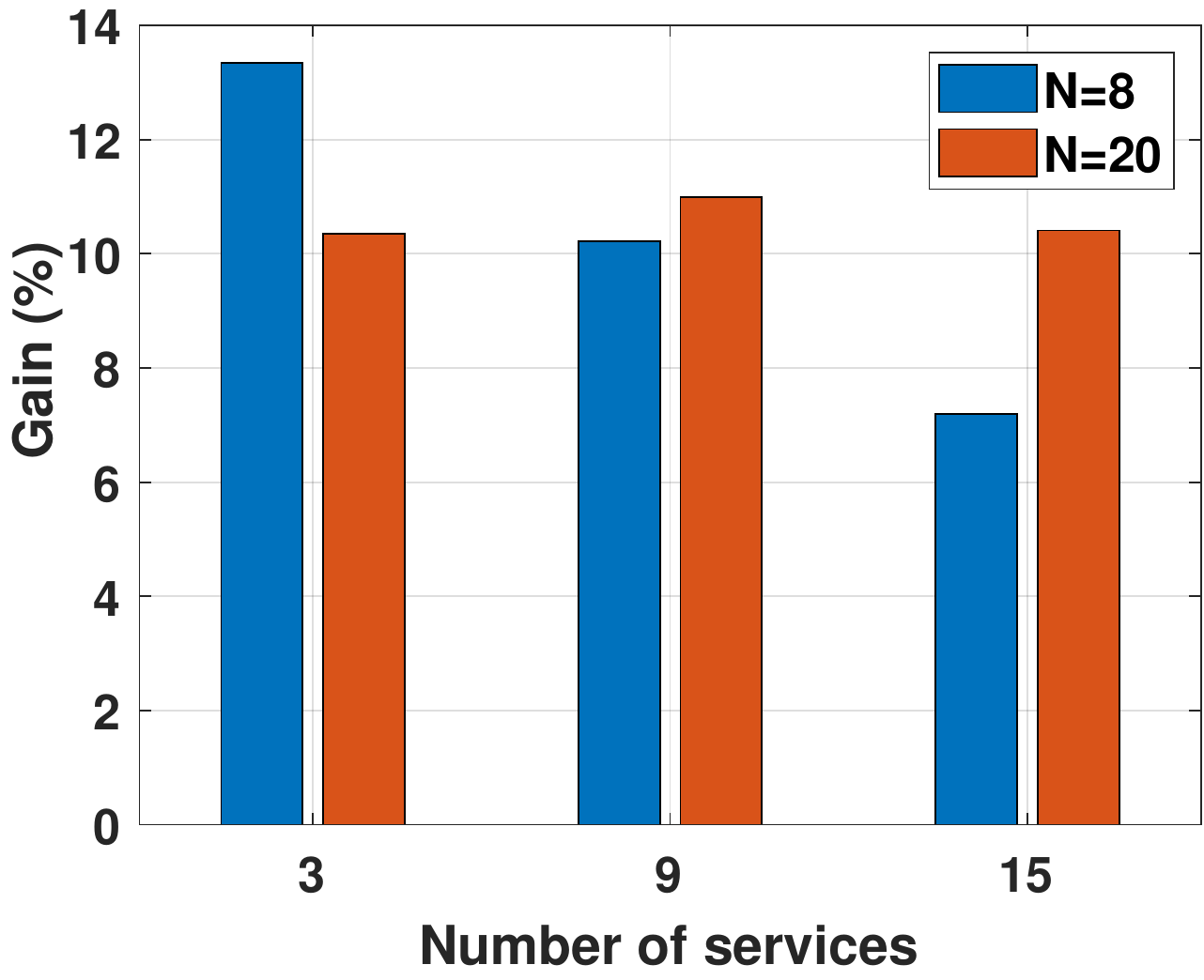}}
\subfigure[Hyper-exponential Service Time]{
\includegraphics[width=0.23\textwidth]{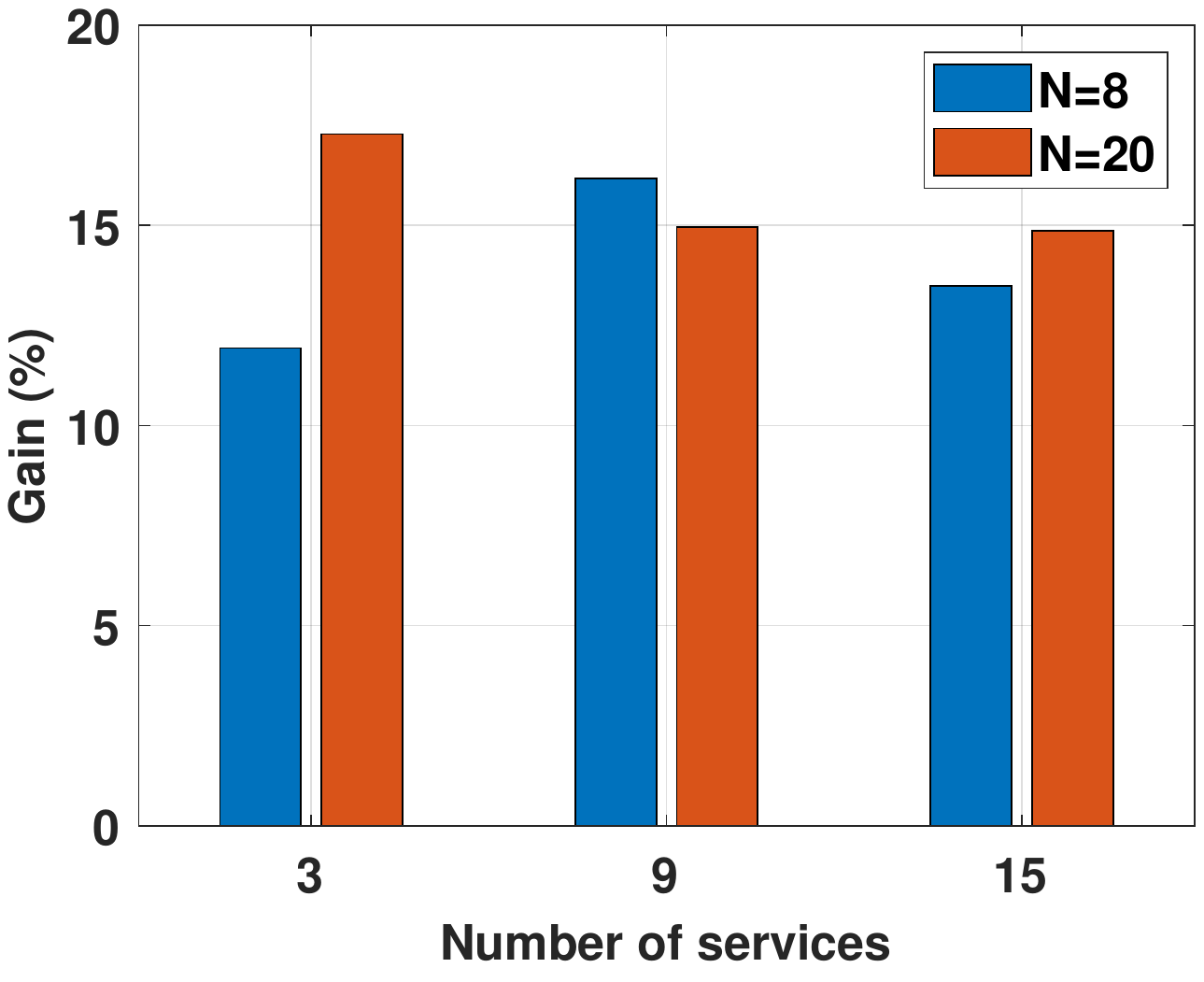}}
\caption{Average gain of response time of the LQN-based model compared to the method in \cite{dependenttask} ($M=2$)}
\label{compareflow1}
\vspace{-0.4cm}
\end{figure}

\begin{figure} [t]
\centering
\subfigure[Exponential Service Time]{
\includegraphics[width=0.23\textwidth]{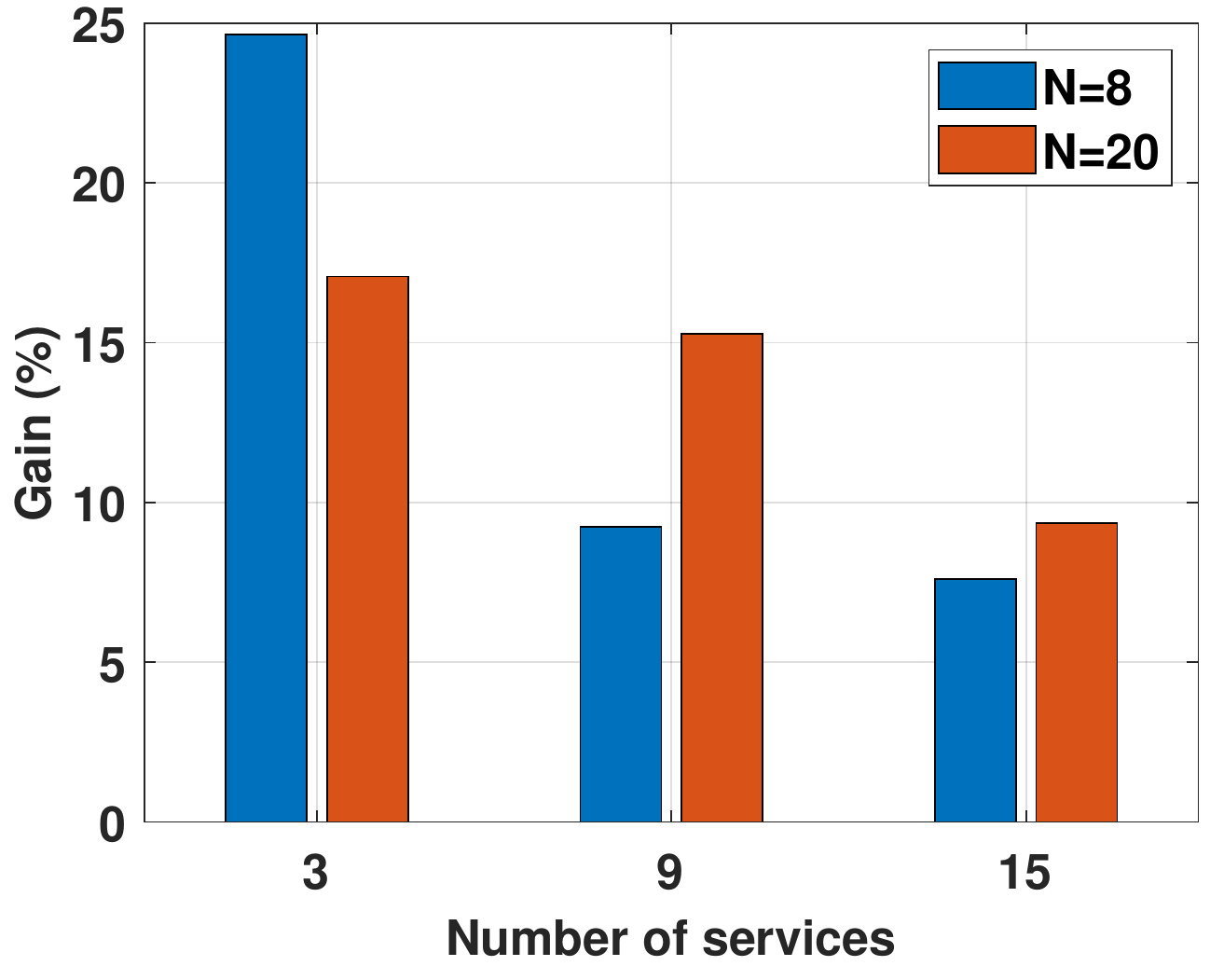}}
\subfigure[Hyper-exponential Service Time]{
\includegraphics[width=0.23\textwidth]{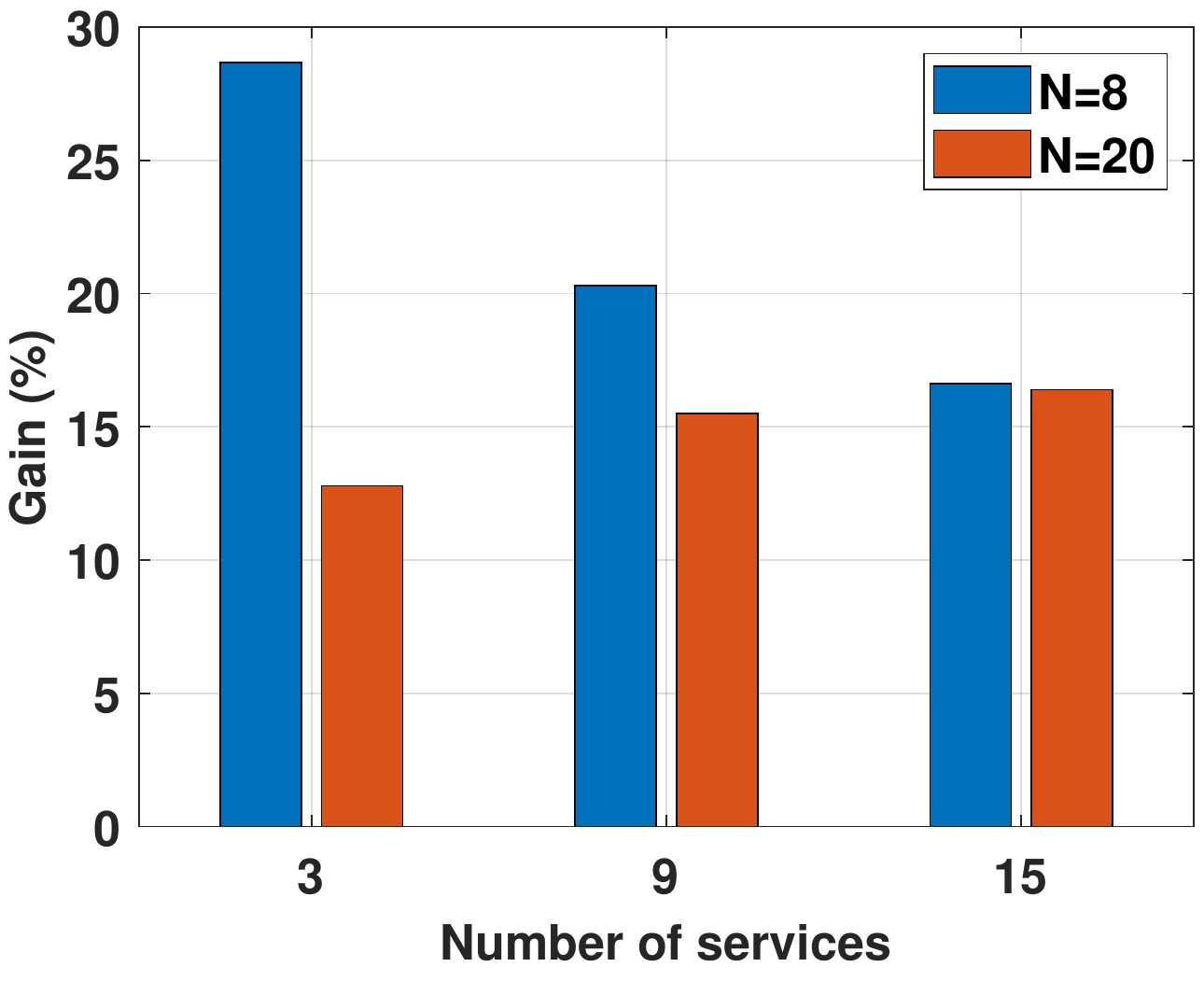}}
\caption{Average gain of response time of the LQN-based model compared to the method in \cite{dependenttask} ($M=5$)}
\label{compareflow2}
\vspace{-0.4cm}
\end{figure}

The experimental results are presented in Fig. \ref{compareflow1} and \ref{compareflow2}, which can be seen that using LQN models yields a better average performance than the method in \cite{dependenttask}. When $M=2$, the gain of total response time is a minimum of 7\%, an average of 12\%, and a maximum of 17\%.  When $M=5$, the gain of total response time is a minimum of 7\%, an average of 16\%, and a maximum of 28\%. As a whole, the average gain improves with the increase of the number of edge nodes. Our results show that it is beneficial to apply LQNs to stochastic scheduling as opposed to heuristically apply a deterministic scheduling, \emph{i.e.,} the solution to the ODT-SC problem. Therefore, it is meaningful to use the LQN model for further performance analysis of edge caching systems.

\subsection{Generalized LQN Model Validation}
We validate the generalized LQN model by JMT, which supports the simulation of queueing networks and Petri nets with a graphical user interface. Petri nets allow us to model the replacement policies in caches as synchronized transitions that alter the cache slots. Thus, by simulating a QPN-based model, we can get an accurate estimate of the expected performance of an LQN with caching, against which we compare our FPI-based implementation described in Algorithm 1.

We consider a validation model containing the upper two layers of the model shown in Fig. \ref{cachetask}. In the upper layer, the underlying processor $P_1$ adopts the processor sharing strategy and the number of users is represented by the multiplicity of the reference task $RT_1$. In the lower layer, the capacity of the cache-task is set as 1 with a total number of 3 items. All the items obey discrete uniform distribution and take the RR strategy. The underlying processor $P_c$ also employs the processor sharing strategy and the number of tokens in this layer is denoted by the multiplicity of the cache-task.

\begin{figure} [t]
\centering
\includegraphics[width=0.48\textwidth]{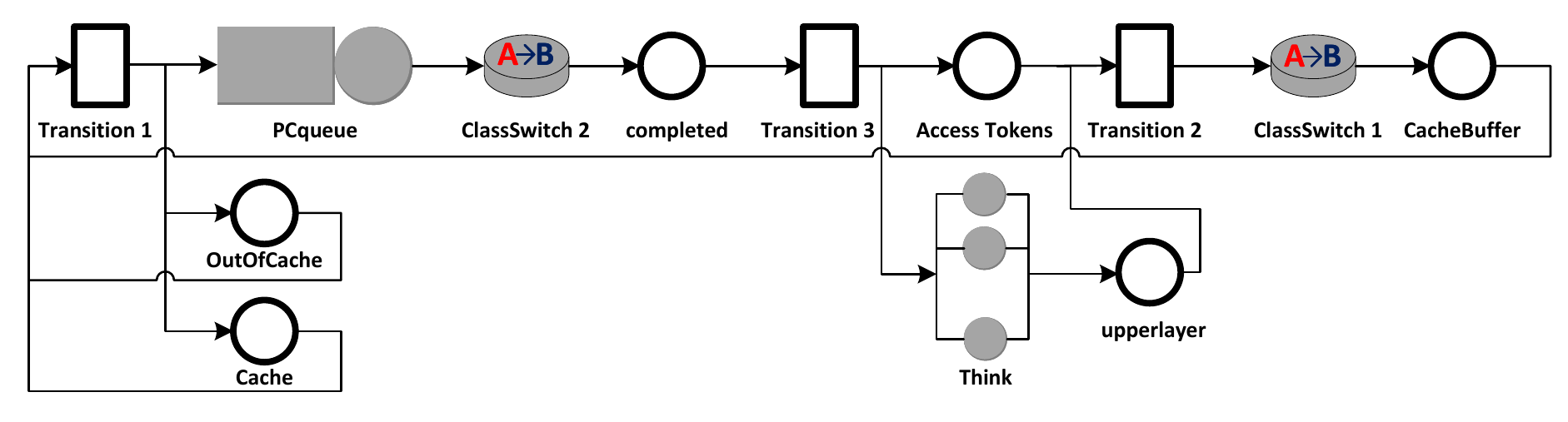}
\caption{A queueing Petri net used for validation via JMT}
\label{jmt}
\end{figure}

The validation model is transformed into a joint framework of queueing networks and Petri nets in JMT, as presented in Fig. \ref{jmt}. The jobs first originate from the delay node and are stored at the place of \emph{upperlayer}. Then, utilizing Petri net transition, \emph{i.e.}, waiting for the tokens to be available in the inbound places and firing a given number of tokens in the output, we enable the transition 2 by 1 job from the upper layer and 1 token, to fire 1 job into the lower layer. To determine a cache hit or miss in the lower layer, we regard the cached items as different job classes. Thus, the fired job class can switch into any of the items at the \emph{ClassSwitch 1} node according to the probability distribution. Based on the state of the stored items in \emph{Cache} and \emph{OutOfCache} places, the transition 1 determines whether to fire a hit or miss job class into the queue. After different services for the hit or miss job at the queue node, the job transforms back to the original job class at the \emph{ClassSwitch 2} node and enables the transition 3. Finally, the transition 3 fires 1 job to the place of \emph{Access Tokens} and 1 job to the delay node to execute iteratively.

\begin{figure} [t]
\centering
\subfigure{
\includegraphics[width=0.23\textwidth]{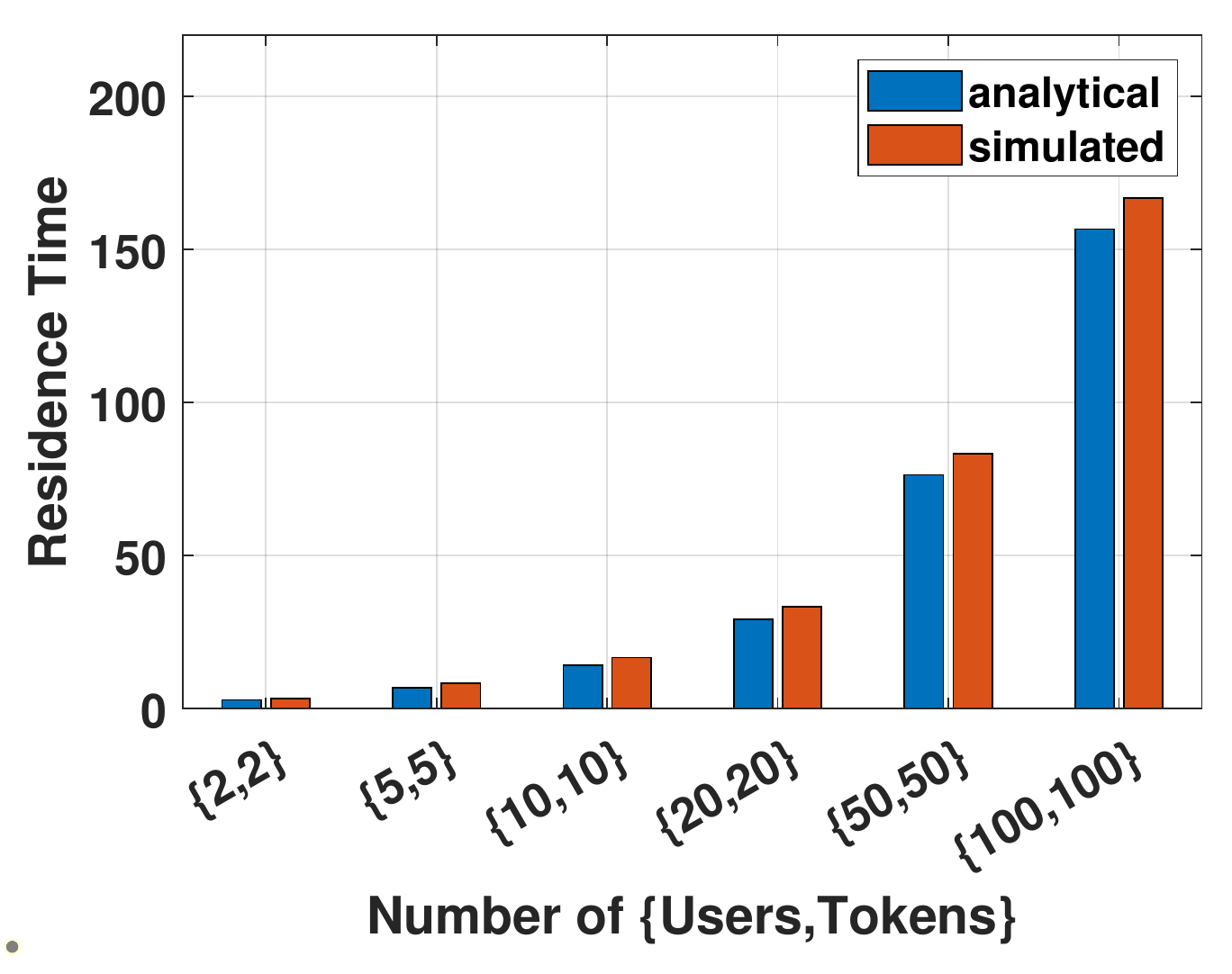}}
\subfigure{
\includegraphics[width=0.23\textwidth]{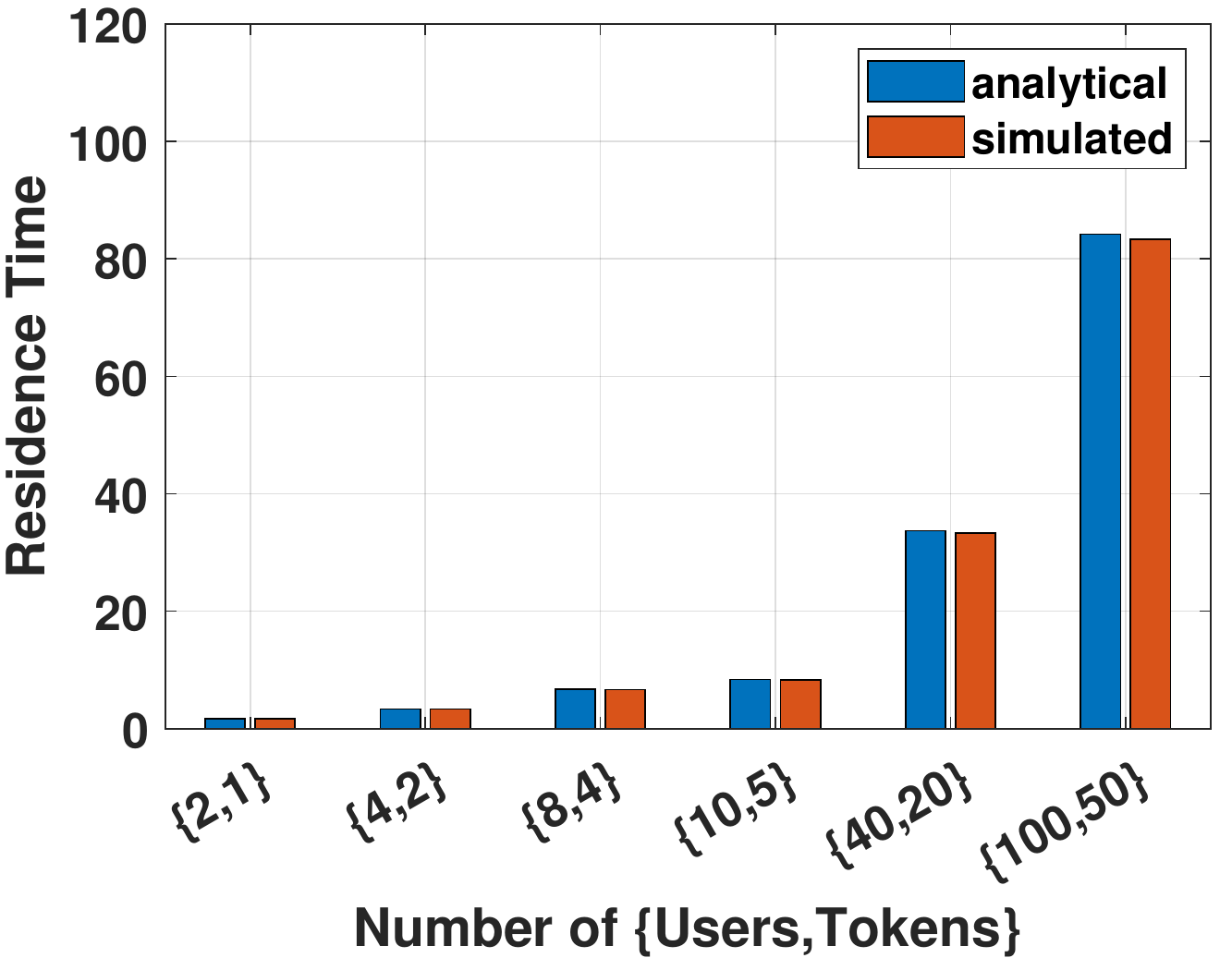}}
\caption{Comparison of residence time between analytical and simulated methods}
\label{jmtresult}
\vspace{-0.4cm}
\end{figure}

We simulate the QPN-based model with JMT to validate the analytical model. The performance metric we select is the residence time, which includes the visit ratio compared to the response time. The results indicate the differences between the analytical and simulated model under different combinations of numbers of users and tokens, as shown in Fig. \ref{jmtresult}. It shows that the differences are negligible in all cases, which proves the high accuracy of the generalized LQN model.

\subsection{Evaluation of the JCSP Method}
To evaluate the JCSP method to real workloads, we perform a trace-driven simulation to demonstrate the system applicability in principle to production systems. The trace is collected from a subset of applications running on Azure Functions in 2019 \cite{azure}. We focus on a random subset of the trace, which contains 34385 application ids and the corresponding serverless function ids belonging to the same application. The invocation rate and the distribution of execution time for each function as well as the distribution of memory usage for each application are also given in the trace.

\begin{figure} [t]
\centering
\subfigure[CDF of Execution Time]{
\includegraphics[width=0.23\textwidth]{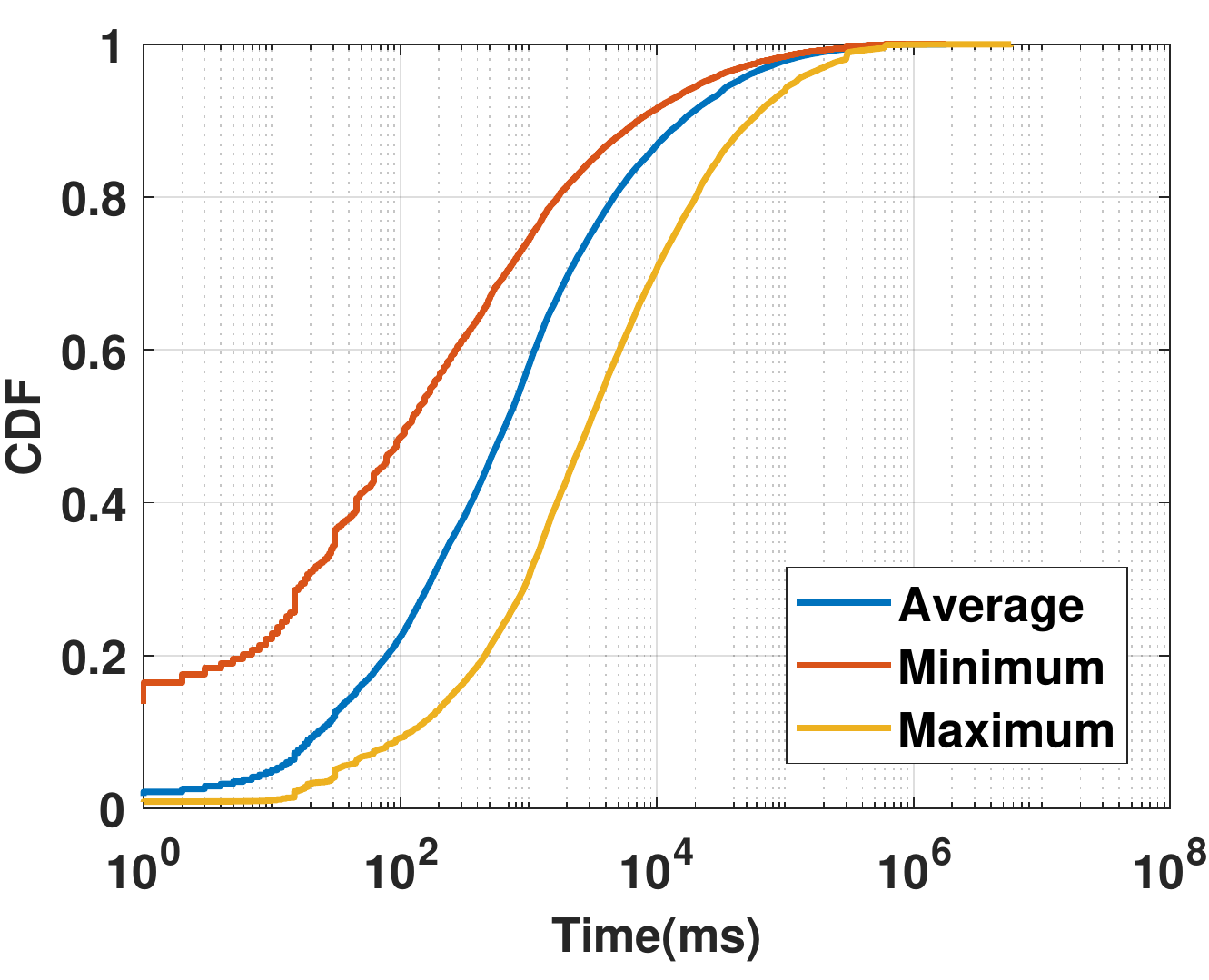}}
\subfigure[CDF of Allocated Memory]{
\includegraphics[width=0.23\textwidth]{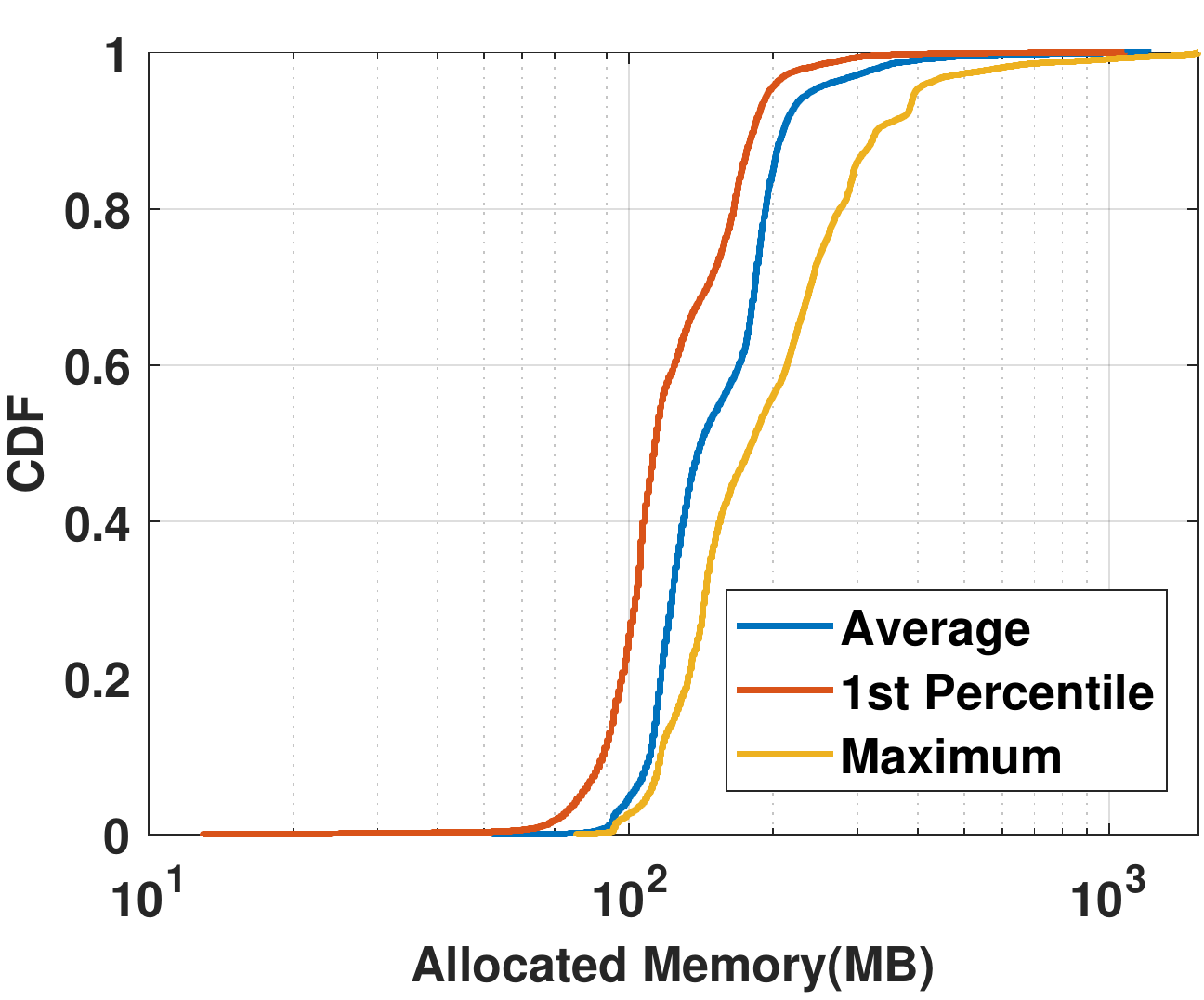}}
\caption{Properties of the selected subset of the real-world Azure traces}
\label{azuredata}
\end{figure}

Fig. \ref{azuredata}(a) shows the cumulative distribution of the minimum, average, and maximum execution time for all functions. We can see that 50\% of the functions have average and maximum execution time less than $0.7s$ and $3s$. 90\% of the functions execute at most $50s$ and 95\% of the functions take less than $50s$ on average. Fig. \ref{azuredata}(b) presents the cumulative distribution of the 1st percentile, average, and maximum virtual memory reserved for all applications. Each application calls for one or more functions. We can observe that 99\% of the applications consume no more than 400MB on average and 99\% of the applications are allocated at most 1000MB.

\begin{figure} [t]
\centering
\includegraphics[width=0.45\textwidth]{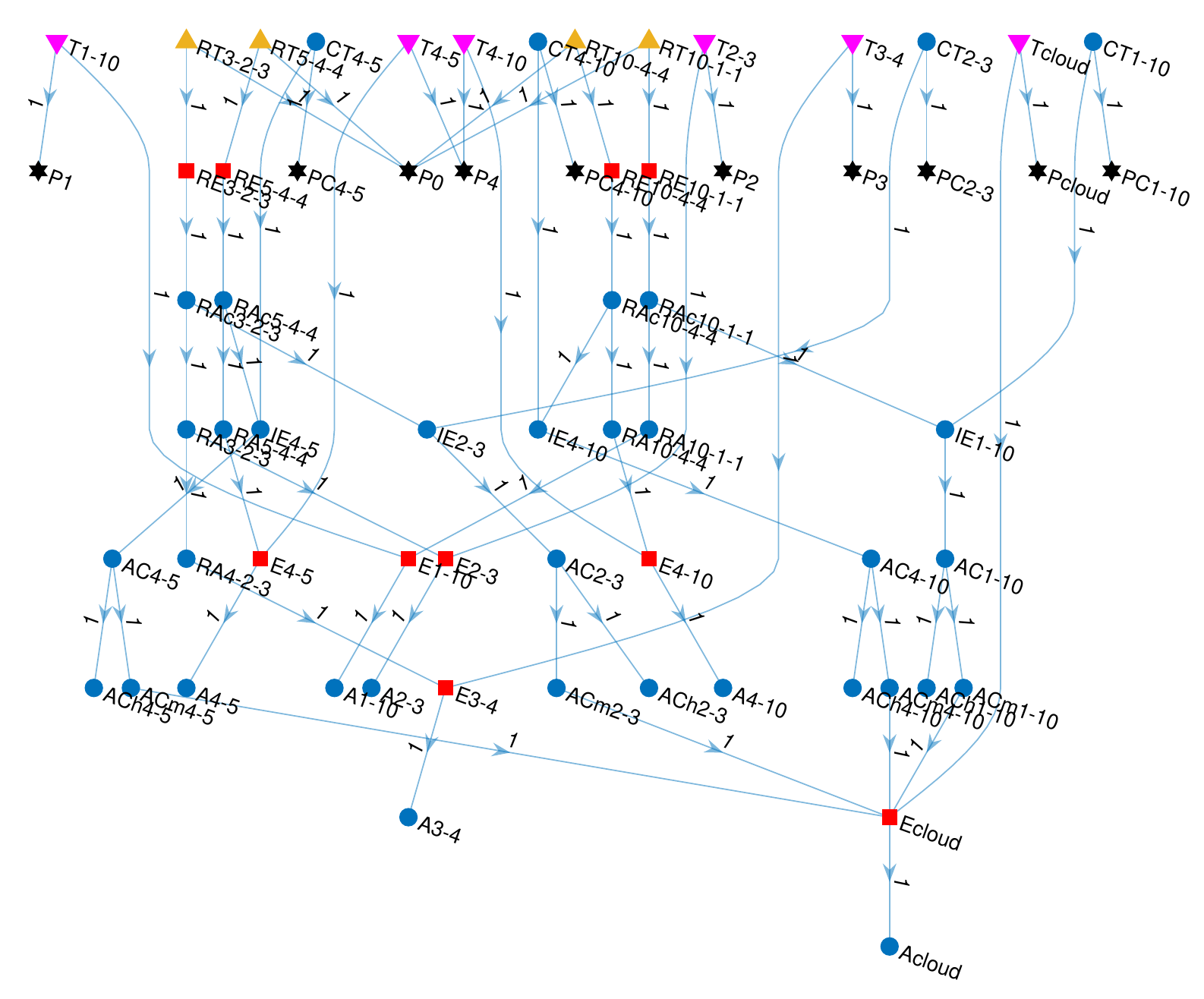}
\caption{{ An example of the generalized LQN model generated by LINE}  }
\label{lineexample}
\vspace{-0.4cm}
\end{figure}

To simulate the trace-driven data, we implement the LQN with caching by LINE \cite{line} to facilitate the JCSP method. An example of the generalized LQN model generated by LINE is shown in Fig. \ref{lineexample}. The dark hexagram, pink triangle, red square, and yellow triangle represent the processor, task, entry, and reference task, respectively. Edges represent client-server relationships. Dots whose name contain \emph{C} and \emph{cloud} belong to the cache  and the origin server. Indexes \emph{h} and \emph{m} refer to cache hit and cache miss. The index \emph{i-j} for the non-reference dots denotes service \emph{j} on node \emph{i}. 

The parameter settings that consider the characteristics of caching are presented in Table \ref{evaluationpara}. The cache capacity $m$ for each node is configured according to a standard Redis caching package offered by Microsoft Azure \cite{cachecapa}. Corresponding cached items obey the Zipf distribution with parameter $\eta$ and are assumed of identical size, the total size $n$ of which is proportional to the node cache capacity. 
These cache-related parameters shown in Table \ref{evaluationpara} are chosen as in earlier works \cite{gast,ton}. 
Compared to Table \ref{exparameters}, the number of edge nodes $d$ and number of services $s$ scales up to 16 and 40 respectively. The distribution of service time is assumed as exponential with a mean value same to the average execution time collected from the Azure trace. The number of users is configured considering the capacity of SBSs \cite{picocells}. For each user, the probability to request for a certain service is determined by the invocation rate of the service.

\begin{table}[t]
\setlength{\belowcaptionskip}{0.5cm}
\centering
\caption{Parameter settings for the JCSP method evaluation} 
\renewcommand\arraystretch{1.25}
{\begin{tabular}{lll  }
\toprule 
\textbf{Definition} &\textbf{Parameter} &\textbf{Range of Value} \\
\hline
number of edge nodes& $M$& $2, 4, 8, 16$  \\
number of users &$N$ & $5, 15, 25$\\
number of services &$C$ & $10, 20, 40$  \\
node cache capacity (MB) &$q$ & $250, 750, 1024$  \\
average total size of items (GB) &$p$ & $0.5, 1, 2, 4, 8$  \\
Zipf parameter &$\eta$ & $0.6, 1.0, 1.4$  \\
\bottomrule
\end{tabular}}
\label{evaluationpara}
\end{table}

For each combination of the parameters, we generate 30 models with different random seeds. In each model, different services on each node are added with cache-tasks that specify the size of total items and the allocated cache capacity for each service. The sum of service capacities is equal to the node cache capacity, which guarantees the maximum utilization of the constrained storage for each edge node. We employ the total system response time as the performance measure and compare our JCSP method to the case of \emph{no cache} and \emph{prefetch all}. The \emph{no cache} scheme deploys no caching module that all the jobs need to wait for the requested contents from origin servers before being processed locally. While the \emph{prefetch all} scheme indicates each edge node caches all the items with a full utilization of the local memory.

\begin{figure} [t]
\centering
\subfigure[$M=2, C/M=5, \eta=1$]{
\includegraphics[width=0.23\textwidth]{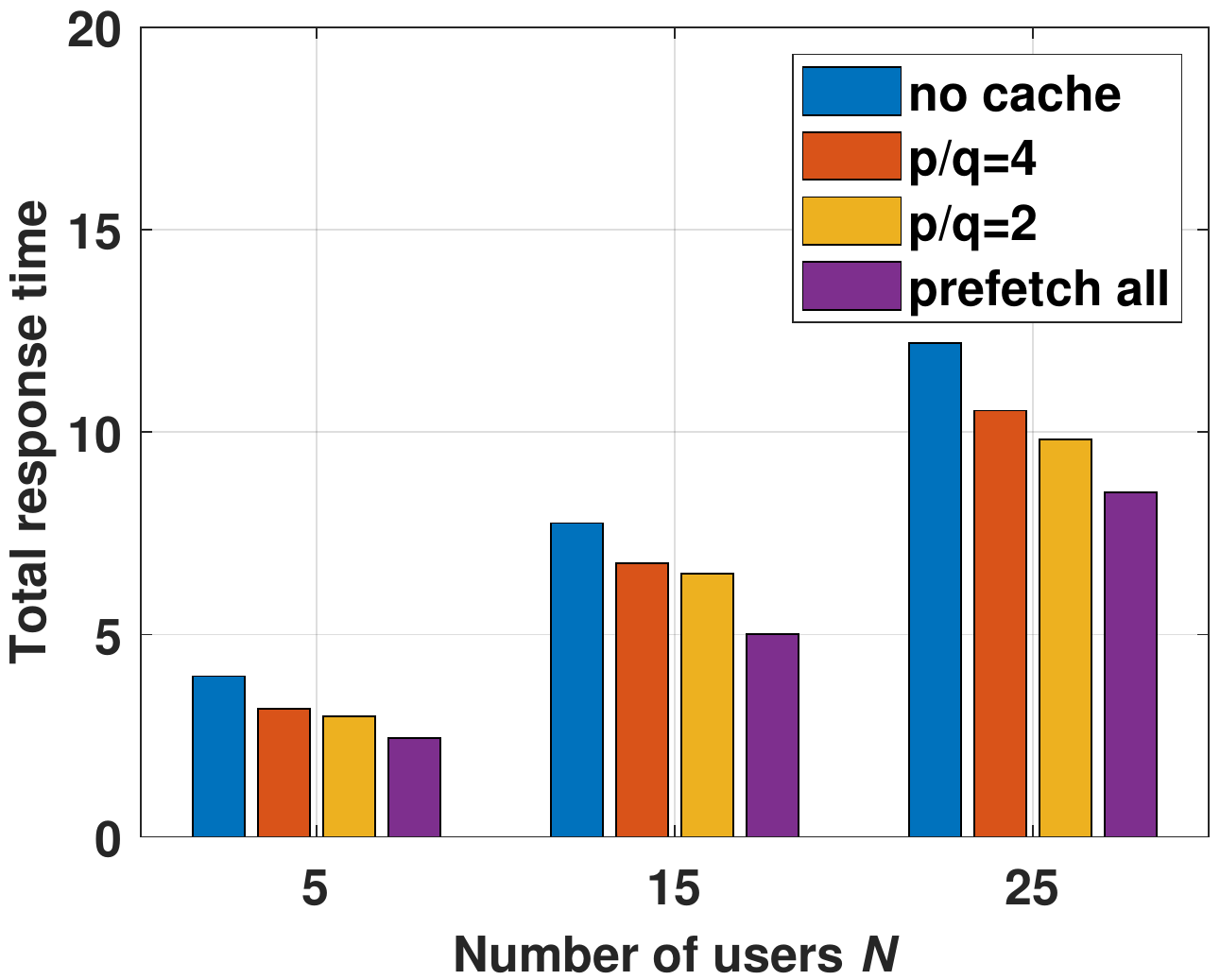}}
\subfigure[$M=4, C/M=5, \eta=1$]{
\includegraphics[width=0.23\textwidth]{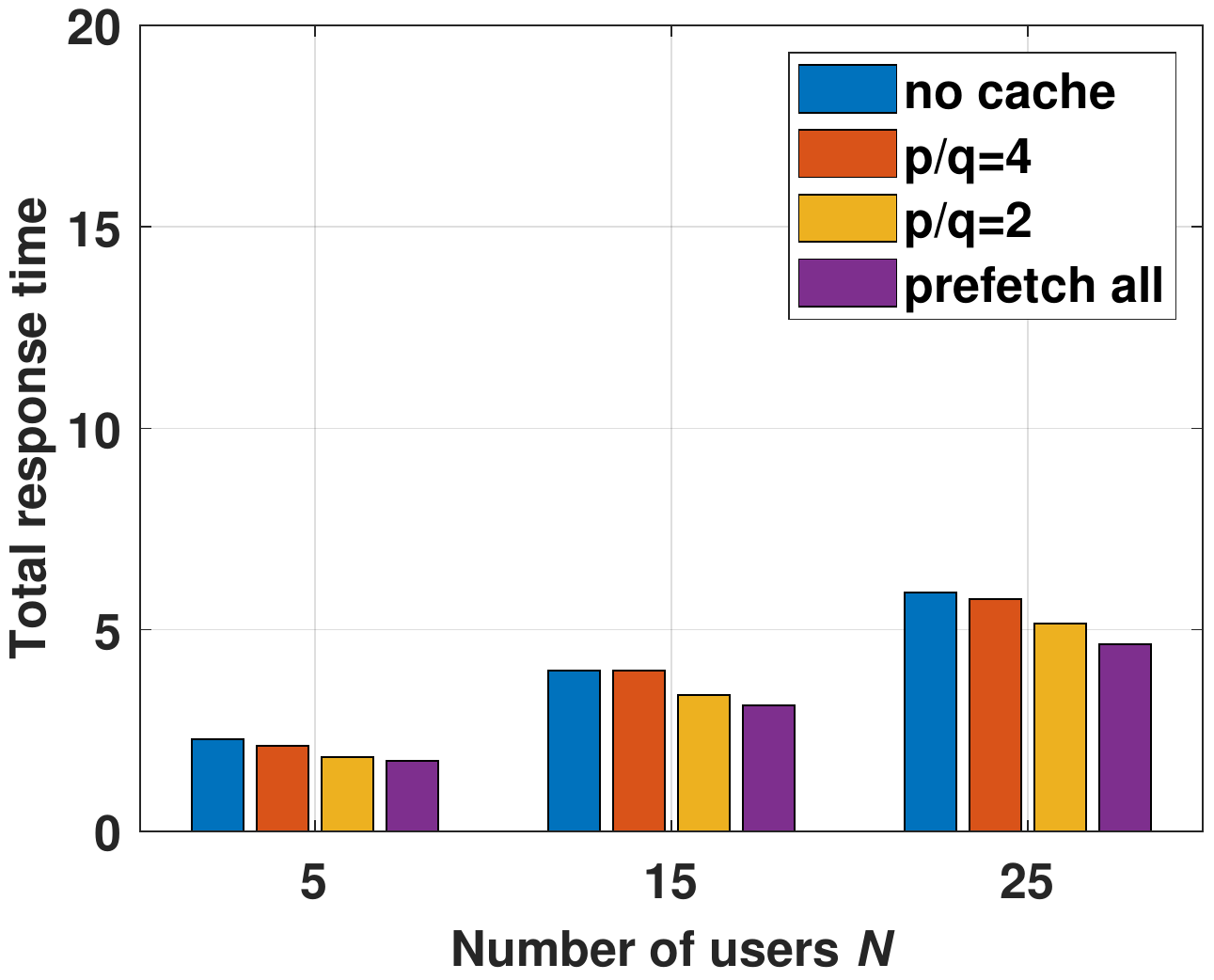}}
\caption{Total response time with respect to the increase of the number of users $N$ (Azure dataset)}
\label{numofusers}
\vspace{-0.4cm}
\end{figure}

From Fig. \ref{numofusers} (a) and (b), we can see that the total response time increases with the growth of the number of users under same conditions, which attributes to the cumulative queueing time. With same number of users, \emph{no cache} scheme as the baseline exhibits the highest response time, because jobs need to obtain required contents from the origin server. Conversely, \emph{prefetch all} scheme exhibits the lowest response time but ignoring the large memory consumption. Compared to the \emph{prefetch all} scheme, although our proposed JCSP method shows an increase of the response time at most 35\% and 30\%, but reduces the memory usage by 500MB and 250MB respectively. This is because our proposed system optimizes the allocation of the cache capacity for each service based on the job scheduling and service placement strategies on different edge nodes, which achieves to maximize the utilization of limited edge resources. Overall, the total response time declines significantly when the number of nodes rises, which is benefit from more concurrent servers that bring down the waiting time.

\begin{figure} [t]
\centering
\subfigure[$M=4, N=25, \eta=1.4$]{
\includegraphics[width=0.23\textwidth]{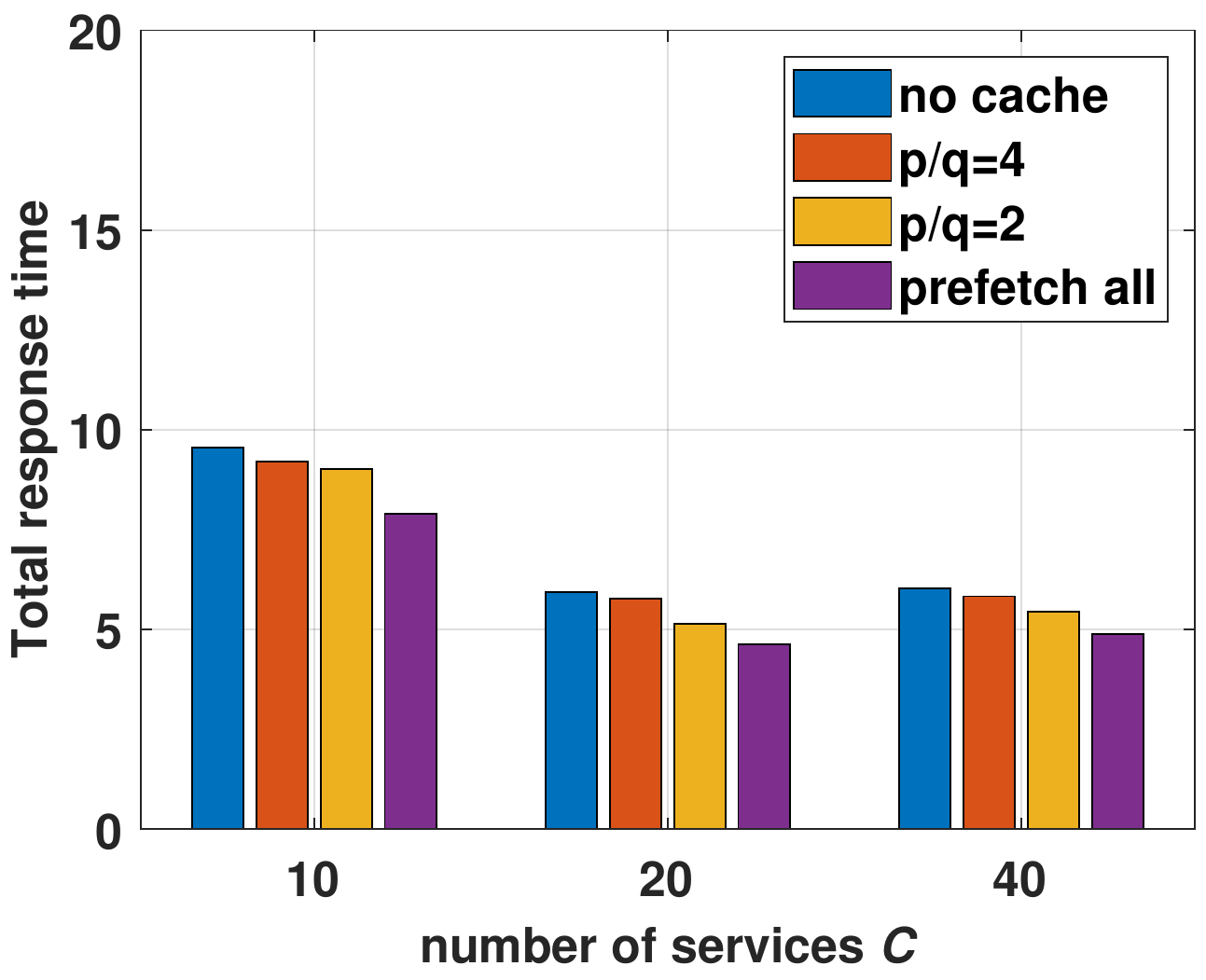}}
\subfigure[$M=8, N=25, \eta=1.4$]{
\includegraphics[width=0.23\textwidth]{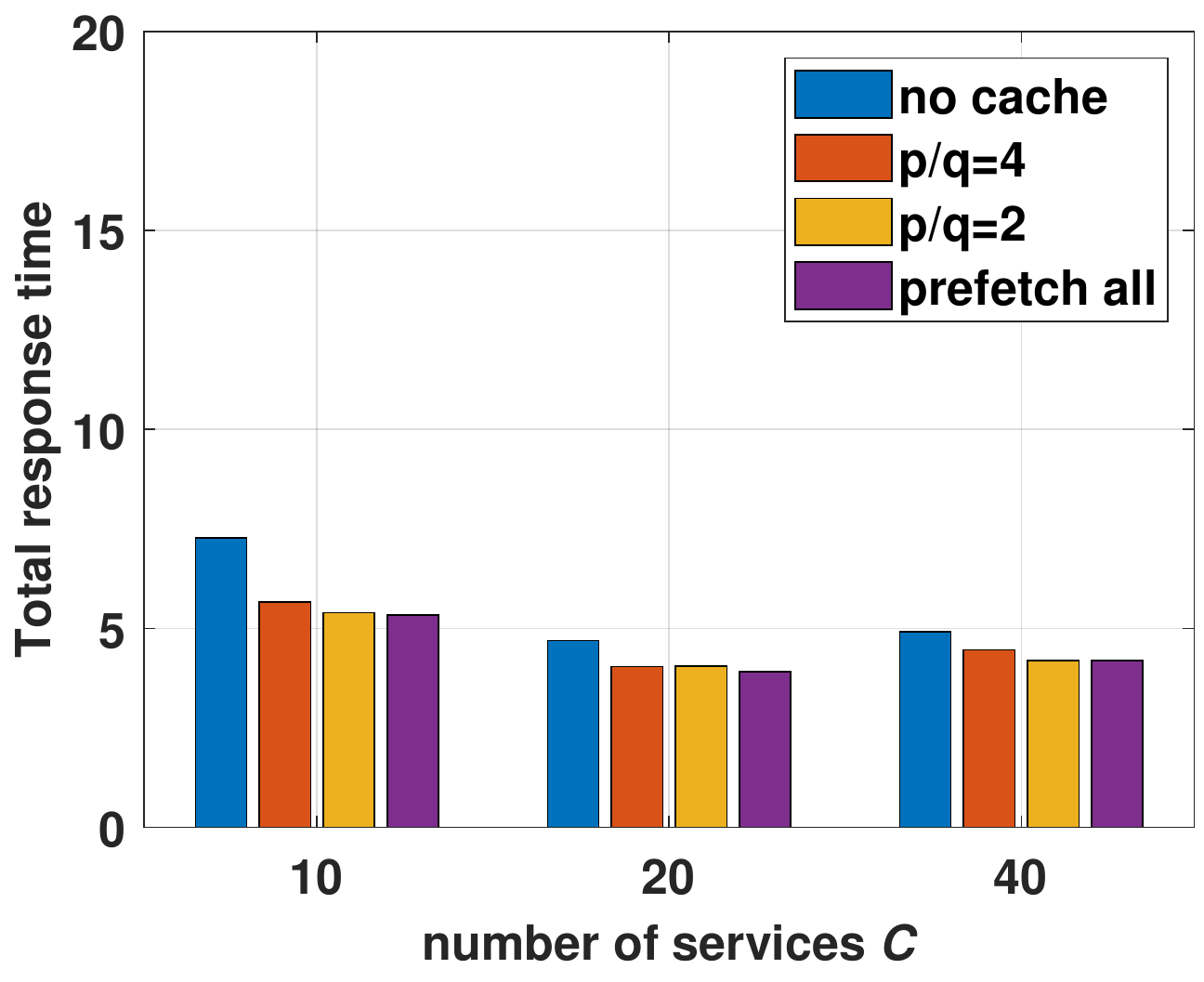}}
\caption{Total response time with respect to the increase of the number of services $C$ (Azure dataset)}
\label{numofservices}
\vspace{-0.4cm}
\end{figure}

From Fig. \ref{numofservices} (a) and (b), we can see that the total response time decreases with the growth of the number of services, but tends to be stable after $C$ reaches 20. The decreasing trend in the first stage is the case when the number of services is less than the number of users. Under the circumstances, the possibility of users to request for the same service is higher, which results in the longer queueing time for the same server. As the number of services grows to equal to or larger than the number of users, the possibility of calling the same service reduces, thus leading to the flat trend in the later stages. With same number of services, our proposed JCSP method only sees an increase of the response time at most 24\% and 14\%, but with a less memory usage of 500MB and 250MB compared to the \emph{prefetch all} scheme. 

\begin{figure} [t]
\centering
\subfigure[$M=8, N=25, C=40$]{
\includegraphics[width=0.23\textwidth]{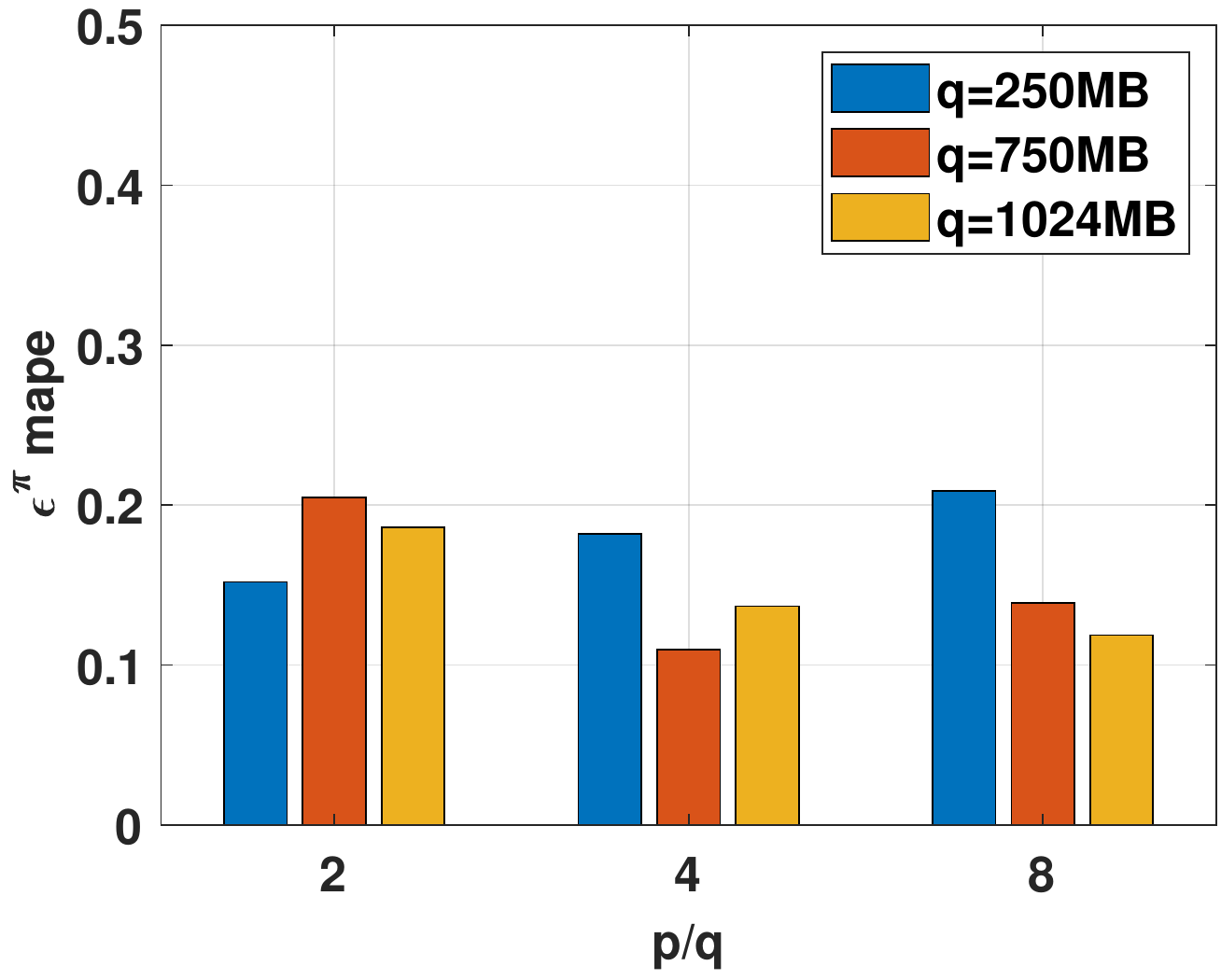}}
\subfigure[$N=25, p=0.5, q=250$]{
\includegraphics[width=0.23\textwidth]{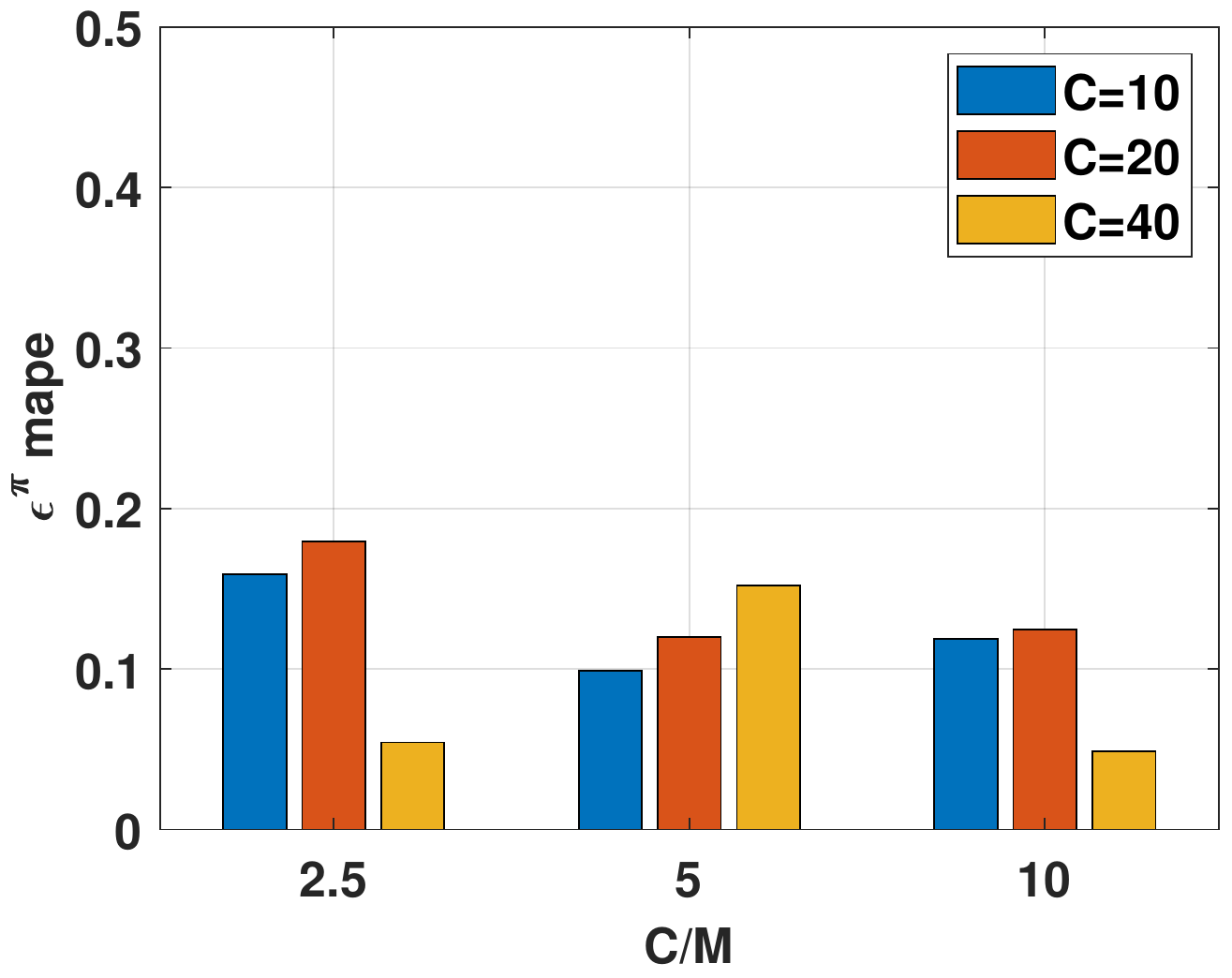}}
\caption{Sensitivity to $p/q$ and $C/M$ ratio with $\eta=1$ (Azure dataset)}
\label{sensitivity}
\vspace{-0.4cm}
\end{figure}

Further, we analyze the sensitivity to $p/q$ and $C/M$ ratio on these experiments. We focus on the miss ratio and define the performance metric as
\begin{equation}
\epsilon^\pi_{mape} =\frac{1}{Q} \frac{1}{M}\frac{1}{C} \sum_{k=1}^Q \sum_{j=1}^M \sum_{i=1}^C \left|1-\frac{\hat{\pi}_{i,j}}{\pi_{i,j}}\right|,
\end{equation}
where $\hat{\pi}_{i,j}$ is the estimated miss ratio for service $i$ at node $j$. The services that are not requested by users on each node are not included. From Fig. \ref{sensitivity} (a), we can see that the average percentage error $\epsilon^\pi_{mape}$ for $p/q$ ratio approximately ranges between $0.1$ and $0.2$. When the node cache capacity is high, the error metric decreases with the size of total items increases. From Fig. \ref{sensitivity} (b), we can observe that the average percentage error $\epsilon^\pi_{mape}$ for $C/M$ ratio approximately ranges between $0.05$ and $0.15$.  

\subsection{Summary of Experimental Results}
Overall, with the increase of the number of users, the JCSP method indicates an increase of the response time at most 35\% and 30\%, but reduces the memory usage by 500MB and 250MB, compared to the \emph{prefetch all} scheme. With the increase of the number of services, the JCSP method shows an increase of the response time at most 24\% and 14\%, but a less memory usage of 500MB and 250MB, compared to the \emph{prefetch all} scheme. This proves our proposed JCSP method is better than simpler heuristics for edge caching resource allocation.

\balance
Concerning time and space requirements for the execution of the JCSP method, the above experiments are conducted on an AMD EPYC 7302P 16-Core Processor. For each replication, the estimate of memory usage and execution time are no more than 4.45MB and $0.2s$, which reflects our proposed JCSP method is scalable for multiple workloads.

\section{Conclusions and Future Work}
In this paper, we present a novel method to jointly analyze edge content caching and service placement. We have first jointly modeled the job scheduling and service placement processes with LQNs. The LQN-based model is compared to the method in \cite{dependenttask} and shows an average gain of 16\% with respect to the response time. Then we have extended the LQN model with caching components to facilitate the JCSP method. An analytical model is given to analyze the generalized LQN model numerically. We further have validated the generalized LQN model by a QPN-based model with JMT and have implemented the proposed class of integrated caching-queueing models within the LINE tool.
Finally, we have conducted extensive trace-driven simulations to evaluate the performance of our proposed JCSP method for a real-world dataset, which show a better trade-off between the response time and memory consumption.

Future work may extend the validation to general phase-type distributions or those involving burstiness (\emph{e.g.,} Markovian arrival processes). Although our implementation supports them, the JCSP method could be further used to understand the impact of splitting a cache into lists on end-to-end edge workflow latency.



\begin{thebibliography}{1}
\addcontentsline{toc}{section}{References}
\bibitem{mobiledevice}
T. Taleb, K. Samdanis, B. Mada, and \emph{et al.}, ``On Multi-Access Edge Computing: A Survey of the Emerging 5G Network Edge Cloud Architecture and Orchestration," \emph{IEEE Commun. Surveys $\&$ Tuts.}, vol. 19, no. 3, pp. 1657-1681, May 2017.

\bibitem{constrain}
L. Lin, X. Liao, H. Jin, and \emph{et al.}, ``Computation Offloading Toward Edge Computing," \emph{Proc. IEEE}, vol. 107, no. 8, pp. 1584-1607, Aug. 2019.

\bibitem{continent}
Y. Mao, C. You, J. Zhang, and \emph{et al.}, ``A Survey on Mobile Edge Computing: The Communication Perspective," \emph{IEEE Commun. Surveys $\&$ Tuts.}, vol. 19, no. 4, pp. 2322-2358, Aug. 2017.

\bibitem{emergence}
M. Satyanarayanan, ``The Emergence of Edge Computing," \emph{Computer}, vol. 50, no. 1, pp. 30-39, Jan. 2017.


\bibitem{codedcache}
M. A. Maddah-Ali and U. Niesen, ``Fundamental Limits of Caching," \emph{IEEE Trans. Inf. Theory}, vol. 60, no. 5, pp. 2856-2867, May 2014.

\bibitem{replacement}
S. Wang, X. Zhang, Y. Zhang, and \emph{et al.}, ``A Survey on Mobile Edge Networks: Convergence of Computing, Caching and Communications," \emph{IEEE Access}, vol. 5, pp. 6757-6779, Mar. 2017.

\bibitem{servicecache}
J. Xu, L. Chen and P. Zhou, ``Joint Service Caching and Task Offloading for Mobile Edge Computing in Dense Networks," in \emph{Proc. of IEEE INFOCOM}, Honolulu, HI, Apr. 2018, pp. 207-215.

\bibitem{franktse}
G. Franks, T. Al-Omari, M. Woodside, and \emph{et al.}, ``Enhanced Modeling and Solution of Layered Queueing Networks," \emph{IEEE Trans. Softw. Eng.}, vol. 35, no. 2, pp. 148-161, Mar.-Apr. 2009.

\bibitem{dependenttask}
G. Zhao, H. Xu, Y. Zhao, and \emph{et al.}, ``Offloading Dependent Tasks in Mobile Edge Computing with Service Caching," in \emph{Proc. of IEEE INFOCOM}, Toronto, ON, Canada, Jul. 2020, pp. 1997-2006.


\bibitem{line}
G. Casale, ``Integrated Performance Evaluation of Extended Queueing Network Models with LINE,” in \emph{Proc. of WSC}, 2020, pp. 2377–2388.

\bibitem{knownpopu}
K. Shanmugam, N. Golrezaei, A. G. Dimakis, and \emph{et al.}, ``FemtoCaching: Wireless Content Delivery Through Distributed Caching Helpers," \emph{IEEE Trans. Inf. Theory}, vol. 59, no. 12, pp. 8402-8413, Dec. 2013.

\bibitem{constpopu1}
B. N. Bharath, K. G. Nagananda and H. V. Poor, ``A Learning-Based Approach to Caching in Heterogenous Small Cell Networks," \emph{IEEE Trans. Commun.}, vol. 64, no. 4, pp. 1674-1686, Apr. 2016.

\bibitem{varypopu1}
A. Sengupta, S. Amuru, R. Tandon, and \emph{et al.}, ``Learning Distributed Caching Strategies in Small Cell Networks," in \emph{Proc. of ISWCS}, Barcelona, Spain, Aug. 2014, pp. 917-921.

\bibitem{varypopu2}
P. Yang, N. Zhang, S. Zhang, and \emph{et al.}, ``Content Popularity Prediction Towards Location-Aware Mobile Edge Caching," \emph{IEEE Trans. Multimedia}, vol. 21, no. 4, pp. 915-929, Apr. 2019.

\bibitem{edgecachingicc}
Z. Xie and W. Chen, ``Storage Efficient Edge Caching with Time Domain Buffer Sharing at Base Stations," in \emph{Proc. of IEEE ICC}, Shanghai, China, May 2019, pp. 1-6.

\bibitem{usercentric}
S. Zhang, P. He, K. Suto, and \emph{et al.}, ``Cooperative Edge Caching in User-Centric Clustered Mobile Networks," \emph{IEEE Trans. Mobile Comput.}, vol. 17, no. 8, pp. 1791-1805, Aug. 2018.

\bibitem{proactivecache}
Y. Liu, A. Zhang, X. Xia, and \emph{et al.}, ``Proactive Data Caching and Replacement in the Edge Computing Environment," in \emph{Proc. of IEEE CLOUD}, Beijing, China, Oct. 2020, pp. 193-200.

\bibitem{vr}
L. Wang, L. Jiao, T. He, and \emph{et al.}, ``Service Entity Placement for Social Virtual Reality Applications in Edge Computing," in \emph{Proc. of IEEE INFOCOM}, Honolulu, HI, Apr. 2018, pp. 468-476.

\bibitem{availability}
I. Lera, C. Guerrero and C. Juiz, ``Availability-Aware Service Placement Policy in Fog Computing Based on Graph Partitions," \emph{IEEE Internet Things J.}, vol. 6, no. 2, pp. 3641-3651, Apr. 2019.

\bibitem{collaborate}
Z. Xu, L. Zhou, S. Chi-Kin Chau, and \emph{et al.}, "Collaborate or Separate? Distributed Service Caching in Mobile Edge Clouds," in \emph{Proc. of IEEE INFOCOM}, Toronto, ON, Canada, Jul. 2020, pp. 2066-2075.

\bibitem{loaddispatching}
L. Yang, J. Cao, G. Liang, and \emph{et al.}, ``Cost Aware Service Placement and Load Dispatching in Mobile Cloud Systems," \emph{IEEE Trans Comput.}, vol. 65, no. 5, pp. 1440-1452, May 2016.

\bibitem{servicecach1}
J. Xu, L. Chen and P. Zhou, ``Joint Service Caching and Task Offloading for Mobile Edge Computing in Dense Networks," in \emph{Proc. of IEEE INFOCOM}, Honolulu, HI, USA, Apr. 2018, pp. 207-215.

\bibitem{servicecach2}
X. Ma, A. Zhou, S. Zhang, and \emph{et al.}, ``Cooperative Service Caching and Workload Scheduling in Mobile Edge Computing," in \emph{Proc. of IEEE INFOCOM}, Toronto, ON, Canada, Jul. 2020, pp. 2076-2085.

\bibitem{functionconfig}
L. Liu, H. Tan, S. H.-C. Jiang, and \emph{et al.}, ``Dependent Task Placement and Scheduling with Function Configuration in Edge Computing," in \emph{Proc. of IEEE/ACM IWQoS}, Phoenix, AZ, USA, Jun. 2019, pp. 1-10.

\bibitem{etsi}
Multi-access Edge Computing (MEC); Phase2: Use Cases and Requirements, ETSI Group Spec. MEC 002, V2.1.1, Oct. 2018.

\bibitem{lqn}
G. Franks, T. Al-Omari, M. Woodside, and \emph{et al.}, ``Enhanced Modeling and Solution of Layered Queueing Networks," \emph{IEEE Trans. Softw. Eng.}, vol. 35, no. 2, pp. 148-161, Mar.-Apr. 2009.

\bibitem{facerecognition}
W. Zhao, R. Chellappa, P. J. Phillips, and \emph{et al.}, ``Face
recognition: A literature survey,” \emph{ACM Comput. Surv. (CSUR)},
vol. 35, no. 4, pp. 399–458, 2003.

\bibitem{ttl}
N. C. Fofack, P. Nain, G. Neglia, and \emph{et al.}, ``Performance Evaluation of Hierarchical TTL-based Cache Networks," \emph{Comput. Netw.}, vol. 65, pp. 212-231, Jun. 2014.

\bibitem{ton}
G. Casale and N. Gast, ``Performance Analysis Methods for List-Based Caches With Non-Uniform Access," \emph{IEEE/ACM Trans. Netw.}, vol. 29, no. 2, pp. 651-664, Dec. 2020.

\bibitem{bolch}
G. Bolch, S. Greiner, H. De Meer, and \emph{et al.}, \emph{Queueing networks and Markov chains}, John Wiley \& Sons, 2006.

\bibitem{gast}
N. Gast and B. V. Houdt, ``Transient and Steady-state Regime of a Family
of List-based Cache Replacement Algorithms,” in \emph{Proc. of SIGMETRICS},
pp. 123–136, ACM, 2015.

\bibitem{jmt}
M. Bertoli, G. Casale and G. Serazzi, ``JMT: Performance Engineering Tools for System Modeling,” \emph{ACM SIGMETRICS PER}, vol. 36, no. 4,
pp. 10–15, 2009.

\bibitem{azure}
M. Shahrad, R. Fonseca, I. Goiri, and \emph{et al.}, ``Serverless in the Wild: Characterizing and Optimizing the Serverless Workload at a Large Cloud Provider," in \emph{Proc. of USENIX ATC 20}, Jul. 2020, pp. 205-218.

\bibitem{cachecapa}
D. T. Hoang, D. Niyato, D. N. Nguyen, and \emph{et al.}, ``A Dynamic Edge Caching Framework for Mobile 5G Networks," \emph{IEEE Wireless Commun.}, vol. 25, no. 5, pp. 95-103, Oct. 2018.

\bibitem{picocells}
``A Guide to 5G Small Cells and Macrocells," https://www. essentracomponents.com/en-gb/news/guides/guide-to-5g-small-cells-and-macrocells, Accessed: 2021-12-31.


\end{thebibliography}
\end{document}